\documentclass[prd,preprint,%superscriptaddress,
tightenlines,nofootinbib,
  eqsecnum%,showpacs
  ]{revtex4}

\usepackage{amsmath}
\usepackage{amsfonts}
\usepackage{amssymb}
\usepackage{bm}
\usepackage{hyperref}
\usepackage{mathrsfs}
\usepackage{graphicx}

\usepackage{ulem}
\usepackage[usenames]{color}

\definecolor{darkgreen}{rgb}{0,0.5,0}

\hypersetup{
    bookmarks=true,         % show bookmarks bar?
    unicode=false,          % non-Latin characters in Acrobat???s bookmarks
    pdftoolbar=true,        % show Acrobat???s toolbar?
    pdfmenubar=true,        % show Acrobat???s menu?
    pdffitwindow=false,     % window fit to page when opened
    pdfstartview={FitH},    % fits the width of the page to the window
    pdftitle={My title},    % title
    pdfauthor={Author},     % author
    pdfsubject={Subject},   % subject of the document
    pdfcreator={Creator},   % creator of the document
    pdfproducer={Producer}, % producer of the document
    pdfkeywords={keyword1} {key2} {key3}, % list of keywords
    pdfnewwindow=true,      % links in new window
    colorlinks=true,       % false: boxed links; true: colored links
    linkcolor=red,          % color of internal links
    citecolor=cyan,        % color of links to bibliography
    filecolor=magenta,      % color of file links
    urlcolor=darkgreen,           % color of external links
    linktocpage=true
}

\allowdisplaybreaks
\DeclareSymbolFontAlphabet{\mathrsfs}{rsfs}
\DeclareMathAlphabet{\mathcal}{OMS}{cmsy}{m}{n}

\newcommand{\scri}{\mathrsfs{I}} 
\newcommand{\ud}{\mathrm{d}}

\newcommand{\beq}{\begin{equation}}
\newcommand{\eeq}{\end{equation}}

\newcounter{theorem} 

\begin{document}

\title{Flux-balance equations for linear momentum and center-of-mass position
  of self-gravitating post-Newtonian systems}

\author{Luc \textsc{Blanchet}}\email{luc.blanchet@iap.fr}
\affiliation{$\mathcal{G}\mathbb{R}\varepsilon{\mathbb{C}}\mathcal{O}$,
  Institut d'Astrophysique de Paris,\\ UMR 7095, CNRS, Sorbonne
  Universit{\'e},\\ 98\textsuperscript{bis} boulevard Arago, 75014 Paris,
  France}

\author{Guillaume \textsc{Faye}}\email{guillaume.faye@iap.fr}
\affiliation{$\mathcal{G}\mathbb{R}\varepsilon{\mathbb{C}}\mathcal{O}$,
  Institut d'Astrophysique de Paris,\\ UMR 7095, CNRS, Sorbonne
  Universit{\'e},\\ 98\textsuperscript{bis} boulevard Arago, 75014 Paris,
  France}

\date{\today}

\begin{abstract}
  We revisit the problem of flux-balance equations for isolated post-Newtonian
  matter systems due to the emission of gravitational waves. In particular we
  show by a local derivation confined to the system, using the expression of
  radiation-reaction forces up to the 3.5PN order, that not only the energy,
  angular momentum and linear momentum of the system, but also the position of
  its center of mass, obey some (non-trivial) flux-balance equations. The
  balance equation for the center-of-mass position completes the description
  of the secular evolution by gravitational waves of relativistic
  post-Newtonian isolated matter systems. We then confirm this result by a
  direct computation of the gravitational-wave fluxes at future null infinity,
  obtaining the full multipole-moment expansion of the flux associated with
  the center-of-mass position (probably new with this paper), and rederiving
  as well the known multipole-moment expansions of the fluxes of energy,
  angular momentum and linear momentum. We also check our analysis by a direct
  calculation of radiation-reaction effects in the case of compact binary
  systems.
\end{abstract}

%\pacs{04.25.Nx, 04.30.-w, 97.60.Jd, 97.60.Lf}

\maketitle

\section{Introduction and motivation} 
\label{sec:intro}

Flux-balance equations and gravitational radiation reaction play a
major role in gravitational physics since the discovery of the binary
pulsar in 1974~\cite{HulseTaylor}. At that time, the
``radiation-reaction controversy''~\cite{Ehletal76,WalkW80} aimed at
reconciling the effect of radiation-reaction
forces~\cite{CE70,Bu71,BuTh70,K80a,K80b,PapaL81} on the orbit of the
binary pulsar with the expected gravitational-wave fluxes for energy
and angular momentum~\cite{E18,LL,PM63,Pe64,EH75,Wag75}. Since then
the controversy was resolved~\cite{Dhouches,D83} and we know now that
the flux-balance equations are correct, at least for extended fluid
systems in an approximate ``post-Newtonian''
sense~\cite{BD88,BD89,BD92,B93,B97}.\footnote{Following the usual
  post-Newtonian terminology, a term $\sim (v/c)^n$ is said to be of
  order $n$PN. Keep in mind that there is always a difference of 2.5PN
  order between the equations of motion and the gravitational
  radiation field, which corresponds to the order $(v/c)^5$ of
  radiation-reaction forces.} Nowadays, the balance equations are used
for building accurate gravitational-wave templates for the data
analysis of inspiralling compact binaries, since they permit to
compute the evolution of the orbital phase and frequency of
inspiralling compact binaries prior their final merger, \textit{i.e.},
the famous ``chirp'' of gravitational
waves~\cite{BlanchetLR,BuonSathya15}.

At leading order, corresponding either to 2.5PN order in the equations
of motion or Newtonian order in the radiation field, the energy
balance is given by the Einstein quadrupole formula~\cite{E18,LL}
\begin{equation}\label{balanceE}
\frac{\ud E}{\ud t} = - \frac{G}{5 c^5} \!\!\stackrel{(3)}{I}_{\!\!ij}
  \stackrel{(3)}{I}_{\!\!ij} 
  + \mathcal{O}\left( \frac{1}{c^7}\right) \,,
\end{equation}
where $I_{ij}$ is the mass-type quadrupole moment of the source, the
superscript $(n)$ denotes time derivatives, and $E$ in the left-hand
side (LHS) is the total mechanical energy of the system. At this order
$E$ is just the constant rest-mass energy plus the Newtonian
mechanical energy. However, there is also a contribution to $E$ at the
2.5PN order, which comes from the right-hand side (RHS) of the balance
equation, when one decides to write the energy flux into ``canonical''
form, \textit{i.e.}, the form of the familiar Einstein quadrupole
formula. Indeed, a total time derivative appears, in the process of
transforming the RHS, and is transferred to the LHS of the balance
equation. This derivative represents the analogue of the Schott term
in electromagnetism~\cite{Schott}. Note that while the flux in the RHS
of~\eqref{balanceE} is gauge invariant, the Schott term depends on the
assumed gauge used to express the radiation-reaction
force~\cite{IW93,IW95}. As is well known, such 2.5PN term being in the
form of a total time derivative, is negligible with respect to the
flux entering the RHS in the adiabatic approximation, \textit{i.e.},
when averaged over an orbital scale much smaller than the
radiation-reaction time scale. The balance equation for angular
momentum reads similarly at the leading order~\cite{Papa71,Th80}
\begin{equation}\label{balanceJ}
\frac{\ud J_i}{\ud t} = - \frac{2G}{5c^5} \, \varepsilon_{ijk}
\!\!\stackrel{(2)}{I}_{\!\!jl}\stackrel{(3)}{I}_{\!\!kl} +
\mathcal{O}\left( \frac{1}{c^7}\right) \,,
\end{equation}
where $J_{ij}$ is the current-type quadrupole moment of the
source. Again there is in the LHS a total time derivative (Schott like
term) at 2.5PN order which is taken into account by a definition of
the angular momentum $J_i$. The flux-balance equations for energy and
angular momentum~\eqref{balanceE}--\eqref{balanceJ} can be combined to
compute the secular evolution (on a radiation-reaction time scale) of
eccentric compact binaries~\cite{PM63,Pe64}.

Concerning the linear momentum, the effect is subdominant as it
appears at order 3.5PN rather than 2.5PN. Indeed, the leading
radiation-reaction force at 2.5PN order integrates to zero over the
system and there is no net force at that order. Including the
radiation reaction at the next 3.5PN order, one obtains the balance
equation for linear momentum as~\cite{BoR61,Peres62,Papa71,Bek73,Th80}
\begin{equation}\label{balanceP}
\frac{\ud P_i}{\ud t} = - \frac{G}{c^7}\left[ \frac{2}{63}
  \!\!\stackrel{(4)}{I}_{\!\!ijk}\stackrel{(3)}{I}_{\!\!jk} + \frac{16}{45}
  \varepsilon_{ijk} \!\!\stackrel{(3)}{I}_{\!\!jl}\stackrel{(3)}{J}_{\!\!kl} 
\right] + \mathcal{O}\left(\frac{1}{c^9}\right)\,,
\end{equation}
where the RHS results from a coupling between the mass octupole moment
$I_{ijk}$ and the mass quadrupole, as well as a coupling between the
mass and current quadrupoles. The main application of the flux-balance
equation for linear momentum is the estimation of the total recoil of
the system by gravitational waves. The recoil velocity (or kick) of
the black hole formed by the merger of two black holes has been
computed by post-Newtonian
methods~\cite{Fit83,Wi92,K95,BQW05,RBK09,LBW10}, perturbation
methods~\cite{OoNaka83,FitDet84,NakaOo87,FHH04}, as well as using
numerical relativity~\cite{Camp05,Bak06b,Camp07,GHS18}.

The quantities $E$, $J_i$ and $P_i$ in
Eqs.~\eqref{balanceE}--\eqref{balanceP} can be referred to as the
``Bondi like'' quantities~\cite{BBM62,Sachs62}, that secularly evolve
by gravitational-radiation emission. By integrating the fluxes in the
RHS and transfering these to the LHS, one can obtain the conserved
invariants for the total matter system $+$ gravitational waves and
which can be referred to as ``ADM like'' quantities~\cite{ADM}. For a
conservative system (for instance, stationary), there are ten
Noetherian conserved invariants associated with the ten symmetries of
the Poincar\'e group. In addition to $E$, $J_i$ and $P_i$, there is
also the initial ``position'' of the center of mass, say $Z_i = G_i -
P_i\,t$, where $G_i$ denotes the position of the center of mass
multiplied by the total (constant) mass. The conservation of $Z_i$ is
due to the invariance of the dynamics under Lorentz boosts. For
self-gravitating systems, $G_i$ coincides with the mass-type dipole
moment $I_i$. When gravitational-wave emission is turned on, the
energy, angular momentum and linear momentum obey the balance or
evolution equations~\eqref{balanceE}--\eqref{balanceP}, and we expect
that the center-of-mass position should also obey a similar evolution
equation.

In the present paper we prove that the flux-balance equation for the
center-of-mass position is, as for~\eqref{balanceP}, a subdominant
3.5PN effect, and involves in the flux just a coupling between the
mass octupole moment and the mass quadrupole, without contribution
from the current moment at the lowest order. We find
\begin{equation}\label{balanceG}
\frac{\ud G_i}{\ud t} = P_i - \frac{2G}{21 c^7}
\!\!\stackrel{(3)}{I}_{\!\!ijk}\stackrel{(3)}{I}_{\!\!jk} +
\mathcal{O}\left(\frac{1}{c^9}\right)\,.
\end{equation}
Together
with~\eqref{balanceE}--\eqref{balanceP}, the law~\eqref{balanceG}
completes the description of the secular evolution by gravitational
waves of relativistic isolated systems. It can be rephrased in an
equivalent way by saying that the initial position $K_i$ of the center
of mass obeys
\begin{equation}\label{balanceK}
\frac{\ud K_i}{\ud t} + t\,\frac{\ud P_i}{\ud t} = -\frac{2G}{21 c^7}
\!\!\stackrel{(3)}{I}_{\!\!ijk}\stackrel{(3)}{I}_{\!\!jk} +
\mathcal{O}\left(\frac{1}{c^9}\right)\,.
\end{equation}
The formulas~\eqref{balanceG} or~\eqref{balanceK} do not appear in standard
text-books on general relativity and gravitational waves such
as~\cite{LL,MTW,Weinberg,Wald,Maggiore,PoissonWill}, nor in review articles
like~\cite{Th300,D300,Dcargese,BlanchetLR,BuonSathya15}. An integral
expression of the relevant flux was, however, already mentioned as a
particular term in the mass dipole moment obtained by matching between the
near zone and the far zone (see Eqs.~(3.46) in~\cite{B93} and (2.15)
in~\cite{B97}). Moreover, a flux-balance equation for the center of mass is
given in general form by Eq.~(6.35) of~\cite{PoissonWill}, and the flux of
center-of-mass position has been computed numerically in~\cite{HSW16}.

In a recent paper, Kozameh \textit{et al.}~\cite{KNQ18} (see also~\cite{KQ16})
investigated the flux-balance equations for general isolated sources based on
the asymptotic properties of the radiation field and Weyl scalars. They
obtained notably the flux of mass-type dipole moment, equivalent to our
center-of-mass position. From Eqs.~(31) and (41)--(44) in~\cite{KNQ18}, we see
that the leading-order contribution to the flux as found there is in
complete agreement with our result~\eqref{balanceG}, modulo a total time
derivative. Furthermore, Ref.~\cite{KNQ18} gives the next-to-leading-order
contribution, involving the current type quadrupole and octupole moments,
which is also in agreement with the corresponding term in the general
multipole expansion obtained in Sec.~\ref{sec:flux}.

In another interesting recent work, Nichols~\cite{N18} derived
related expressions for the flux of center-of-mass position,
which he refers to as the CM angular momentum but is really, like for
us, the Noetherian invariant associated with the boost symmetry of the system.
The calculations in~\cite{N18} are performed at future null infinity using the
Bondi-Sachs formalism in the general case, as well as in the leading PN
approximation, and are motivated by the study of the memory effect induced by
changes in the multipole moments parametrizing the CM angular momentum. More
work should be done in order to compare the results of the present paper to
those of Ref.~\cite{N18}.

The fluxes of energy, angular momentum and linear momentum are known
as full multipole-moment series expansions, formally extending
Eqs.~\eqref{balanceE}--\eqref{balanceP} up to any multipolar
order~\cite{Th80}. The multipole expansions are given in terms of the
so-called ``radiative'' multipole moments, which parametrize the
waveform at infinity. These multipole expansions are thus exact by
definition, although the radiative moments are not directly connected
to the source. In the present paper, we also perform the flux
computations and rederive the known multipole expansions for energy,
angular momentum and linear momentum up to any multipolar order, but
we adopt a different strategy and express the fluxes in terms of
appropriate ``source'' multipole moments that are known as explicit
integrals over the matter source. The prize we have to pay is that our
calculation is valid only at the leading post-Minkowskian order,
\textit{i.e.}, the dominant order in $G$. By the same method, we then
obtain the complete multipole expansion of the flux associated with
the position of the center of mass, thus generalizing
Eq.~\eqref{balanceG} to all multipolar contributions in terms of the
source multipole moments.

The plan of this paper is as follows. In Sec.~\ref{sec:radreac}, we
recall from previous works the expression in a specific gauge of the
radiation-reaction force for extended isolated systems to 3.5PN
order. In Sec.~\ref{sec:balance} we use that expression to derive the
flux-balance equations for all invariant quantities and, in
particular, that for the center-of-mass position to the lowest order,
proving Eq.~\eqref{balanceG}. In Sec.~\ref{sec:flux}, we perform a
direct flux calculation at future null infinity for all the
invariants, yielding the multipole expansions generalizing the lowest
order results~\eqref{balanceE}--\eqref{balanceG} to any multipolar
order (at dominant order in $G$). Finally in Sec.~\ref{sec:eom}, we
reconfirm the results for the linear momentum and center-of-mass
position by working out the 3.5PN harmonic-coordinate
radiation-reaction force in the case of compact binary systems. The
paper ends in Sec.~\ref{sec:disc} with a discussion on the meaning of
the results and a short conclusion.

\section{Gravitational radiation reaction to 3.5PN order} 
\label{sec:radreac}

The first derivation we propose is valid for a general isolated
(compact-support) post-Newtonian matter system. It is based on a
specific expression of the radiation-reaction force at the 3.5PN
order, defined in a particular gauge which is an appropriate extension
of the Burke-Thorne gauge~\cite{Bu71,BuTh70} for the lowest-order
radiation reaction. The 1PN radiation reaction in this gauge has been
obtained in~\cite{B93,B97}. In fact, the present calculation will be
the continuation of the paper~\cite{B97}. The radiation-reaction
effects are entirely specified once we give the metric, which admits
the following components, accurate to order 3.5PN concerning
radiation-reaction (dissipative) effects:
\begin{subequations}\label{metric}
\begin{align}
  g_{00} &= -1 + \frac{2\mathcal{V}}{c^2} - \frac{2\mathcal{V}^2}{c^4}
  + \frac{1}{c^6} \mathop{g}_{6}{}_{\!\!00} + \frac{1}{c^8}
  \mathop{g}_{8}{}_{\!\!00} +
  \mathcal{O}\left(\frac{1}{c^{10}}\right)\,,\\ g_{0i} &=
  -\frac{4\mathcal{V}_i}{c^3} + \frac{1}{c^5}
  \mathop{g}_{5}{}_{\!\!0i} + \frac{1}{c^7} \mathop{g}_{7}{}_{\!\!0i}
  + \mathcal{O}\left(\frac{1}{c^{9}}\right)\,,\\ g_{ij} &= \delta_{ij}
  \left(1 + \frac{2\mathcal{V}}{c^2}\right) + \frac{4}{c^4}\Bigl(
  W_{ij} - \delta_{ij} W_{kk} \Bigr) + \frac{1}{c^6}
  \mathop{g}_{6}{}_{\!\!ij} +
  \mathcal{O}\left(\frac{1}{c^{8}}\right)\,.
\end{align}
\end{subequations}
Concerning conservative effects, this metric is only accurate to 1PN
order, but notice the term of order $1/c^4$ in the spatial metric
$g_{ij}$, which is included for completeness and is really of order
2PN for the motion of massive particles and 1PN for the motion of
photons. This term is important to control the energy of the system at
1PN order or, equivalently, the total mass up to 2PN order,
see~\cite{B97}. The coefficients ${}_ng_{\mu\nu}$ in~\eqref{metric}
represent some 2PN and 3PN \textit{conservative} effects, uncontrolled
at this stage, but which will not contribute to the present
calculation. The metric is characterized by a scalar potential
$\mathcal{V}$, a vector potential $\mathcal{V}_i$, and a tensorial one
$W_{ij}$. These potentials are generated by the matter stress-energy
tensor $T^{\mu\nu}$ of the compact-support source through the
effective mass, current and stress densities:
\begin{equation}\label{sigma}
  \sigma = \frac{T^{00} + T^{ii}}{c^2} \,,\qquad \sigma_i =
  \frac{T^{0i}}{c} \,,\qquad \sigma_{ij} = T^{ij} \,,
\end{equation}
with $T^{ii}=\delta_{ij}T^{ij}$. The potentials $\mathcal{V}_\mu =
(\mathcal{V}, \mathcal{V}_i)$ contain a conservative part and a
radiation-reaction part, hence we pose
\begin{equation}\label{Vmu}
\mathcal{V}_\mu = V^\text{sym}_\mu + V^\text{reac}_\mu \,.
\end{equation}
For the present purpose, the conservative part is \emph{defined} by
the ``symmetric'' propagator acting on the matter source proportional
to $\sigma_\mu=(\sigma, \sigma_i)$, \textit{i.e.},
\begin{equation}\label{sym}
  V^\text{sym}_\mu = \Box^{-1}_\text{sym}\bigl(-4 \pi \,\sigma_\mu\bigr) =
  \sum_{k=0}^{+\infty} \left(\frac{\partial}{c\partial
      t}\right)^{2k}\Delta^{-k-1}\bigl(-4 \pi \,\sigma_\mu\bigr)\,.
\end{equation}
Strictly speaking, such definition does not yield a purely
conservative quantity since the matter stress-energy
tensor~\eqref{sigma} depends on the metric~\eqref{metric} and, thus,
does contain some radiation-reaction contributions. With this caveat
in mind, we still keep this definition since it is very convenient, as
we shall see. To an order consistent with~\eqref{metric}, we can
express the previous ``symmetric'' potentials as
\begin{subequations}\label{Ucons}
\begin{align}
  V^\text{sym} &= U + \frac{1}{2c^2}\partial_t^2X + \frac{1}{c^4}
  \mathop{V}_{4}{}^\text{sym} + \frac{1}{c^6}
  \mathop{V}_{6}{}^\text{sym} +
  \mathcal{O}\left(\frac{1}{c^{8}}\right)\,,\\ V_i^\text{sym} &= U_i +
  \frac{1}{c^2} \mathop{V}_{2}{}_{\!\!i}^\text{sym} + \frac{1}{c^4}
  \mathop{V}_{4}{}_{\!\!i}^\text{sym} +
  \mathcal{O}\left(\frac{1}{c^{6}}\right)\,,
\end{align}
\end{subequations}
where we do not need to control the 2PN and 3PN terms while we have
introduced the usual Poisson integrals $U_\mu$ and the
``super-potential'' $X$ (only for the scalar potential) defined by:
\begin{equation}\label{potUX}
  U_\mu = G \int
  \frac{\ud^3\mathbf{x}'}{\vert\mathbf{x}-\mathbf{x}'\vert}
  \,\sigma_\mu(\mathbf{x}',t)\,,\qquad X = G \int
  \ud^3\mathbf{x}'\,\vert\mathbf{x}-\mathbf{x}'\vert
  \,\sigma(\mathbf{x}',t) \,.
\end{equation}
As for the tensorial part $W_{ij}$ in the metric~\eqref{metric}, it is
given (modulo some constant factor) by the Poisson integral of the
matter stresses $\sigma_{ij}$ plus the usual Newtonian gravitational
stresses:
\begin{equation}\label{Wij}
  W_{ij} = G \int \frac{\ud^3\mathbf{x}'}{\vert\mathbf{x}-\mathbf{x}'\vert}
  \left[\sigma_{ij} + \frac{1}{4\pi G}\Bigl(\partial_i U\partial_j U -
    \frac{1}{2} \delta_{ij} \partial_k U \partial_k U\Bigr)\right]\,.
\end{equation}
The radiation-reaction effects in $W_{ij}$ are negligible within the present
approximation.

All the information regarding radiation reaction is ``implicitly''
contained in the matter currents $\sigma_\mu$ (as we have seen), and
``explicitly'' into the scalar and vector radiation-reaction
potentials $V_\mu^\text{reac}=(V^\text{reac},V_i^\text{reac})$. Those
are defined in terms of the multipole moments $I_L(t)$ and $J_L(t)$ of
the post-Newtonian source as~\cite{B93,B97}\footnote{The multipole
  moments are symmetric and trace-free (STF) with respect to their
  $\ell$ indices $L=i_1i_2\cdots i_\ell$; we denote the STF product of
  spatial vectors $x_i$ as $\hat{x}_L=\text{STF}(x_L)$, so that
  $\hat{x}_{ij}=x_{ij}-\frac{1}{3}\delta_{ij}r^2$,
  $\hat{x}_{ijk}=x_{ijk}-\frac{1}{5}(\delta_{ij}x_k+\delta_{ik}x_j +
  \delta_{jk}x_i)r^2$ and so on (with $r=\vert\mathbf{x}\vert$,
  $x_{ij}=x_ix_j$ and $x_{ijk}=x_ix_jx_k$); similarly, we denote
  $\partial_L=\partial_{i_1}\partial_{i_1}\cdots\partial_{i_\ell}$ and
  $\hat{\partial}_L=\text{STF}(\partial_L)$; $\varepsilon_{ijk}$ is
  the fully antisymmetric Levi-Cevita symbol. Notice that all products
  of spatial vectors in Eqs.~\eqref{reac} are actually STF. On the
  other hand, parenthesis surrounding indices mean their
  symmetrization: $t_{(ij)}=\frac{1}{2}(t_{ij}+t_{ji})$ whereas
  superscripts $(n)$ stand for $n$ time derivatives.}
\begin{subequations}\label{reac}
\begin{align}
V^\text{reac} &= - \frac{G}{5 c^5} \,x_{ij}
\!\!\stackrel{(5)}{I}_{\!\!ij} +
\frac{G}{c^7}\left[\frac{1}{189}x_{ijk}\!\!\stackrel{(7)}{I}_{\!\!ijk}
  - \frac{1}{70} r^2 x_{ij} \!\!\stackrel{(7)}{I}_{\!\!ij} \right] +
\mathcal{O}\left(\frac{1}{c^{8}}\right)\,,\\ V_i^\text{reac} &=
\frac{G}{c^5}\left[\frac{1}{21}\hat{x}_{ijk}
  \!\!\stackrel{(6)}{I}_{\!\!jk} - \frac{4}{45} \varepsilon_{ijk}
  \,x_{jl} \!\!\stackrel{(5)}{J}_{\!\!kl}\right] +
\mathcal{O}\left(\frac{1}{c^{7}}\right)\,.
\end{align}
\end{subequations}
At the 4PN order the scalar potential $V^\text{reac}$ contains the
contribution of tails~\cite{B97}, which is not needed here [but see
  Eq.~\eqref{balancetail} below]. Moreover, at the 4.5PN order, there
should also be a genuine tensorial contribution to the
radiation-reaction force. When restricted, by contrast, to the leading
2.5PN order, the potential $V^\text{reac}$ reduces to the Burke-Thorne
radiation-reaction scalar potential~\cite{Bu71,BuTh70}. The multipole
moments are the mass quadrupole moment $I_{ij}$, consistently given
here with 1PN accuracy~\cite{BD89},
\begin{equation}\label{Iij}
I_{ij} = \int \ud^3\mathbf{x} \left[ \hat{x}_{ij} \sigma +
  \frac{1}{14c^2} r^2 \hat{x}_{ij} \partial_t^2\sigma -
  \frac{20}{21c^2} \,\hat{x}_{ijk} \,\partial_t \sigma_k \right] +
\mathcal{O}\left(\frac{1}{c^{4}}\right)\,,
\end{equation}
the mass octupole moment $I_{ijk}$ and the current quadrupole
$J_{ij}$, which are merely Newtonian:
\begin{equation}\label{IJ}
  I_{ijk} = \int \ud^3 \mathbf{x}\,\hat{x}_{ijk} \,\sigma +
  \mathcal{O}\left(\frac{1}{c^{2}}\right)\,, \qquad J_{ij} = \int \ud^3
  \mathbf{x}\, \varepsilon_{kl(i} \,x_{j) k} \,\sigma_l +
  \mathcal{O}\left(\frac{1}{c^{2}}\right)\,.
\end{equation}

\section{Flux-balance equations to 3.5PN order} 
\label{sec:balance}

The metric~\eqref{metric} is used in the derivation of the
flux-balance equations for linear momentum and (in an extension
of~\cite{B97}) for the position of the center of mass; concerning the
energy and angular momentum, we shall simply restate the results
of~\cite{B97}. The method consists of integrating the equations of
motion over a volume enclosing the compact-support matter distribution
of the source, the equations of motion being here the covariant
conservation of the matter stress-energy tensor, $\nabla_\nu
T_\mu^\nu=0$, rewritten in the more convenient way as
\begin{equation}\label{nablaT}
\partial_\nu\Pi^\nu_\mu = \mathcal{F}_\mu\,,
\end{equation}
where we pose $\Pi^\nu_\mu = \sqrt{-g}\,T^\nu_\mu$ and
$\mathcal{F}_\mu=\frac{1}{2}\sqrt{-g}\,\partial_\mu
g_{\rho\sigma}\,T^{\rho\sigma}$, with $T^\nu_\mu =
g_{\mu\rho}T^{\nu\rho}$ and $g=\text{det}(g_{\rho\sigma})$.  By
substituting~\eqref{metric} and using the definitions of the mass,
current and stress densities~\eqref{sigma}, we obtain explicit
expressions containing all radiation effects up to 3.5PN order but in
which the conservative effects at 2PN and 3PN orders are
neglected. The components of the effective force in the RHS read
\begin{subequations}\label{Fmu}
\begin{align}
\mathcal{F}_0 &= \frac{1}{c} \sigma \,\partial_t \mathcal{V} -
\frac{4}{c^3} \sigma_j \partial_t \mathcal{V}_j +
\frac{1}{c^5}\mathop{\mathcal{F}}_{5}{}_{\!\!0} +
\frac{1}{c^7}\mathop{\mathcal{F}}_{7}{}_{\!\!0} +
\mathcal{O}\left(\frac{1}{c^{9}}\right)\,,\\ \mathcal{F}_i &= \sigma
\,\partial_i \mathcal{V} - \frac{4}{c^2} \sigma_j \partial_i
\mathcal{V}_j + \frac{1}{c^4}\mathop{\mathcal{F}}_{4}{}_{\!\!i} +
\frac{1}{c^6}\mathop{\mathcal{F}}_{6}{}_{\!\!i} +
\mathcal{O}\left(\frac{1}{c^{8}}\right)\label{Fi}\,,
\end{align}
\end{subequations}
while those of the effective momentum in the LHS are
\begin{subequations}\label{Pimunu}
\begin{align}
\Pi^0_0 &= -c^2 \sigma + \sigma_{ii} +
\frac{4}{c^2}\bigl(\sigma\,W_{ii} - \sigma_i\,\mathcal{V}_i\bigr) +
\frac{1}{c^4}\mathop{\Pi}_{4}{}^{\!\!0}_{\!\!0}+
\frac{1}{c^6}\mathop{\Pi}_{6}{}^{\!\!0}_{\!\!0} +
\mathcal{O}\left(\frac{1}{c^{8}}\right)\,,\\ \Pi^0_i &= c\,\sigma_i +
\frac{4}{c}\bigl(\sigma_i \mathcal{V} - \sigma \mathcal{V}_i\bigr) +
\frac{1}{c^3}\mathop{\Pi}_{3}{}^{\!\!0}_{\!\!i}+
\frac{1}{c^5}\mathop{\Pi}_{5}{}^{\!\!0}_{\!\!i} +
\mathcal{O}\left(\frac{1}{c^{7}}\right)\,\label{Pi0i},\\ \Pi_0^i &= -
c\,\sigma_i + \frac{4}{c^3}\bigl(\sigma_i W_{jj} - \sigma_{ij}
\mathcal{V}_j\bigr) + \frac{1}{c^5}\mathop{\Pi}_{5}{}^{\!\!i}_{\!\!0}
+ \mathcal{O}\left(\frac{1}{c^{7}}\right)\,,\\ \Pi^i_j &= \sigma_{ij}
+ \frac{4}{c^2}\bigl(\sigma_{ij} \mathcal{V} - \sigma_i
\mathcal{V}_j\bigr) + \frac{1}{c^4}\mathop{\Pi}_{4}{}^{\!\!i}_{\!\!j}
+ \frac{1}{c^6}\mathop{\Pi}_{6}{}^{\!\!i}_{\!\!j}+
\mathcal{O}\left(\frac{1}{c^{8}}\right)\,.
\end{align}
\end{subequations}
The 1PN equation of continuity and the 1PN equation of motion (or
relativistic Euler equation) are explicitly given in the present
formalism by
\begin{subequations}\label{contEuler}
\begin{align}
&\partial_t\sigma + \partial_i\sigma_i =
  \frac{1}{c^2}\Bigl(\partial_t\sigma_{jj} - \sigma
  \partial_t\mathcal{V}\Bigr) +
  \overline{\mathcal{O}}\left(\frac{1}{c^{8}}\right)\,,\label{cont}\\ &\partial_t
  \left[\sigma_i \left( 1 + \frac{4\mathcal{V}}{c^2}\right)\right] +
  \partial_j \left[ \sigma_{ij} \left(
    1+\frac{4\mathcal{V}}{c^2}\right)\right] = \sigma\partial_i
  \mathcal{V}+\frac{4}{c^2} \Bigl[ \sigma\partial_t \mathcal{V}_i +
    \sigma_j \bigl(\partial_j \mathcal{V}_i -\partial_i
    \mathcal{V}_j\bigr) \Bigr] +
  \overline{\mathcal{O}}\left(\frac{1}{c^{8}}\right)\,.\label{Euler}
\end{align}
\end{subequations}
The special notation for the remainder term means that it is correct
regarding radiation-reaction effects up to 3.5PN order included, but
contains some uncontrolled 2PN and 3PN conservative
contributions. Namely, the remainders in~\eqref{contEuler} are of the
type
\begin{equation}\label{notationO}
\overline{\mathcal{O}}\left(\frac{1}{c^{8}}\right) =
\frac{1}{c^4}\mathop{X}_{4} + \frac{1}{c^6}\mathop{X}_{6} +
\mathcal{O}\left(\frac{1}{c^{8}}\right)\,.
\end{equation}
The radiation-reaction parts~\eqref{reac} of the potentials
$\mathcal{V}_\mu$ appear explicitly through the dependence of the
laws~\eqref{contEuler} upon $\mathcal{V}_\mu$, but also implicitly
through the matter currents and stresses, $\sigma_\mu$ and
$\sigma_{ij}$, respectively. We find that only the former will
contribute to the fluxes in the RHS of the balance equations, while
the latter implicit radiation-reaction terms appear only as total time
derivatives in the LHS. In order to obtain the radiation-reaction
contributions in the stress-energy tensor $T^{\mu\nu}$, one must
identify the matter degrees of freedom, independently from the
gravitational field (\textit{i.e.}, the metric). For a perfect fluid
system, the matter degrees of freedom can be chosen to be the
coordinate-velocity field $v^\mu=c u^\mu/u^0$, the specific entropy
$s_*$, and the coordinate density $\rho_* = \sqrt{-g}\,\rho\,u^0$,
where $\rho$ is the conserved scalar density satisfying
$\nabla_\mu(\rho u^\mu)=0$, and $u^0=(-g_{\mu\nu}v^\mu
v^\nu/c^2)^{-1/2}$. In this way, we find for instance
$\sigma^\text{reac}=\rho_* V^\text{reac}/c^2+\mathcal{O}(1/c^9)$,
which is a 3.5PN effect.

\subsection{Linear momentum} 
\label{sec:linmom}

The balance equation for linear momentum has already been derived
in~\cite{B97}, but we repeat here the main steps for completeness. We
first integrate Eq.~\eqref{nablaT} with $\mu=i$ over the
compact-support matter distribution. Using the Gauss law to discard a
total divergence, as well as the expressions~\eqref{Fi}
and~\eqref{Pi0i}, we thus obtain
\begin{equation}\label{baleq1}
\frac{\ud}{\ud t}\left(\int\ud^3\mathbf{x}\left[\sigma_i +
  \frac{4}{c^2}\left(\sigma_i \mathcal{V} - \sigma
  \mathcal{V}_i\right)\right]\right) = \int\ud^3\mathbf{x}\left(\sigma
\partial_i \mathcal{V} -
\frac{4}{c^2}\sigma_j\partial_i\mathcal{V}_j\right) +
\overline{\mathcal{O}}\left(\frac{1}{c^{8}}\right)\,.
\end{equation}
We then split the potentials $\mathcal{V}_\mu$ into symmetric and
reaction parts [see Eq.~\eqref{Vmu}] and perform simple
manipulations. In particular, we use a generalization of the
``action-reaction'' theorem valid at the 1PN order, namely
\begin{equation}\label{actionreac}
  \int\ud^3\mathbf{x}\,\sigma \,\partial_i V^\text{sym} =
  \frac{\ud}{\ud t}\left[\frac{1}{2c^2}\int\ud^3\mathbf{x}\,\sigma
    \,\partial_i \partial_t X\right] +
  \mathcal{O}\left(\frac{1}{c^{4}}\right)\,,
\end{equation}
where $X$ is the super-potential defined in~\eqref{potUX}. This
permits transfering a term in the form of a total time derivative to
the LHS of the equation, yielding
\begin{align}\label{baleq2}
&\frac{\ud}{\ud t}\left(\int\ud^3\mathbf{x}\left[\sigma_i -
    \frac{1}{2c^2}\sigma \partial_i\partial_t X +
    \frac{4}{c^2}\Bigl(\sigma_i V^\text{reac} - \sigma
    V^\text{reac}_i\Bigr)\right] +
  \overline{\mathcal{O}}\left(\frac{1}{c^{8}}\right)\right)
  \nonumber\\ &\qquad\qquad\qquad = \int\ud^3\mathbf{x}\left(\sigma
  \partial_i V^\text{reac} - \frac{4}{c^2}\sigma_j\partial_i
  V^\text{reac}_j\right) + \mathcal{O}\left(\frac{1}{c^{9}}\right)\,.
\end{align}
In this equation, we have also transferred the uncontrolled 2PN and
3PN approximations from the RHS to the LHS, where they now lie inside
the remainder [see our special notation~\eqref{notationO}]. Indeed,
these approximations are conservative so that they must appear in the
form of total time derivatives in our flux-balance equations. This has
been proven long ago for general fluids at the 2PN
approximation~\cite{K80a,K80b}, but we assume here that the same is
true for the 3PN terms, for the conservative terms at the 4PN
approximation (\textit{i.e.}, apart for the tail term), and so on.
This assumption has been fully confirmed by recent works on the
equations of motion of compact binary systems at the 4PN order, which
showed the existence at that order of all the Poincar\'e invariants
(see notably~\cite{BBFM17}, and Sec.~\ref{sec:eom} below).

At this stage, the radiation-reaction potentials appear on both sides
of Eq.~\eqref{baleq2}. In the next step, we replace the reaction
potentials in the RHS by their explicit expressions provided by
Eqs.~\eqref{reac}. A straightforward computation leads then to the
usual flux-balance equation for linear momentum at the leading 3.5PN
order:
\begin{equation}\label{dPidt}
\frac{\ud P_i}{\ud t} = - \frac{G}{c^7}\left[ \frac{2}{63}
  \!\!\stackrel{(4)}{I}_{\!\!ijk}\stackrel{(3)}{I}_{\!\!jk} +
  \frac{16}{45} \varepsilon_{ijk}
  \!\!\stackrel{(3)}{I}_{\!\!jl}\stackrel{(3)}{J}_{\!\!kl} \right] +
\mathcal{O}\left(\frac{1}{c^9}\right)\,,
\end{equation}
where, importantly for the precise meaning of the equation, the linear
momentum $P_i$ in the LHS is explicitly given by
\begin{equation}\label{Pi}
P_i = \int\ud^3\mathbf{x}\left[\sigma_i -
  \frac{1}{2c^2}\sigma \partial_i\partial_t X + \frac{4}{c^2}\Bigl(\sigma_i
  V^\text{reac} - \sigma V^\text{reac}_i\Bigr)\right] + \delta P^\text{reac}_i
+ \overline{\mathcal{O}}\left(\frac{1}{c^{8}}\right)\,.
\end{equation}
Besides the neglected conservative 2PN and 3PN terms included in the
remainder, there appears a term $\delta P^\text{reac}_i$, composed of
2.5PN and 3.5PN approximations, which comes from the total time
derivatives originally present in the RHS. As it will play a major
role in what follows, we display it explicitly:
\begin{align}\label{schottPi}
\delta P^\text{reac}_i &=
\frac{2G}{5c^5}\left(I_{j}\!\!\stackrel{(4)}{I}_{\!\!ij} -
\!\!\stackrel{(1)}{I}_{\!\!j} \stackrel{(3)}{I}_{\!\!ij} \right)
\nonumber\\ & + \frac{G}{c^7}\biggl[ \frac{8}{15} J_{j}J^{(4)}_{ij} -
  \frac{1}{63} \!\!\stackrel{(6)}{I}_{\!\!ijk}I_{jk} +
  \frac{1}{63}\!\!\stackrel{(5)}{I}_{\!\!ijk}
  \stackrel{(1)}{I}_{\!\!jk} - \frac{1}{63}
  \!\!\stackrel{(4)}{I}_{\!\!ijk} \stackrel{(2)}{I}_{\!\!jk} -
  \frac{1}{63} \!\!\stackrel{(3)}{I}_{\!\!ijk}
  \stackrel{(3)}{I}_{\!\!jk} \nonumber\\ &\qquad\qquad + \frac{1}{63}
  \!\!\stackrel{(2)}{I}_{\!\!ijk} \stackrel{(4)}{I}_{\!\!jk} -
  \frac{1}{63}
  \!\!\stackrel{(1)}{I}_{\!\!ijk}\stackrel{(5)}{I}_{\!\!jk} +
  \frac{1}{35} I_{ijk}\!\!\stackrel{(6)}{I}_{\!\!jk}
  \nonumber\\ &\qquad - \frac{8}{45}\varepsilon_{ijk}
  \!\!\stackrel{(4)}{J}_{\!\!jl}\stackrel{(1)}{I}_{\!\!kl} +
  \frac{8}{45}\varepsilon_{ijk}
  \!\!\stackrel{(3)}{J}_{\!\!jl}\stackrel{(2)}{I}_{\!\!kl} +
  \frac{8}{45}\varepsilon_{ijk}
  \!\!\stackrel{(2)}{J}_{\!\!jl}\stackrel{(3)}{I}_{\!\!kl} -
  \frac{8}{45}\varepsilon_{ijk}
  \!\!\stackrel{(1)}{J}_{\!\!jl}\stackrel{(4)}{I}_{\!\!kl}
  \nonumber\\ &\qquad\qquad + \frac{8}{45}\varepsilon_{ijk}
  J_{jl}\!\!\stackrel{(5)}{I}_{\!\!kl} + \frac{2}{5}
  K_{j}\!\!\stackrel{(5)}{I}_{\!\!ij} - \frac{1}{5}
  L_{j}\!\!\stackrel{(5)}{I}_{\!\!ij} + \frac{1}{25}
  N_{j}\!\!\stackrel{(6)}{I}_{\!\!ij} \biggr]\,.
\end{align}
The mass quadrupole moment $I_{ij}$ already enters this expression at
the 2.5PN order and is thus consistently given with 1PN accuracy by
Eq.~\eqref{Iij}. In addition, the leading terms depend on the mass
dipole moment, which is required with the same 1PN accuracy and reads
\begin{equation}\label{Ii}
I_i = \int\ud^3\mathbf{x}\,x_i\left( \sigma +
\frac{1}{c^2}\left[\frac{\sigma U}{2} - \sigma_{jj} \right] \right) +
\mathcal{O}\left(\frac{1}{c^4}\right) \,.
\end{equation}
For the expressions of the other moments in~\eqref{schottPi}, which
are simply Newtonian, one should refer to~\eqref{IJ}. Furthermore,
$\delta P^\text{reac}_i$ contains the conserved Newtonian angular
momentum of the system identified with the current-type dipole moment,
\begin{equation}\label{Ji}
J_{i} = \int \ud^3 \mathbf{x}\, \varepsilon_{ijk} \,x_j \,\sigma_k +
\mathcal{O}\left(\frac{1}{c^2}\right)\,,
\end{equation}
such that $\ud J_i/\ud t=\mathcal{O}(1/c^2)$, as well as three
suplementary integrals defined at Newtonian order only,
\begin{equation}\label{KLMi}
K_{i} = \int \ud^3 \mathbf{x}\, \sigma_j \,x_{ij}\,,\quad L_{i} = \int
\ud^3 \mathbf{x}\, \sigma_i \,r^2 \,,\quad N_{i} = \int \ud^3
\mathbf{x}\, \sigma \,x_i \,r^2 \,.
\end{equation}
These integrals are not independent but are linked together, as a
consequence of the continuity equation~\eqref{cont}, by $N^{(1)}_{i} =
2 K_i + L_i + \mathcal{O}(1/c^2)$.

\subsection{Center-of-mass position} 
\label{sec:CM}

We now want to rewrite the linear momentum $P_i$ defined by
Eqs.~\eqref{Pi}-\eqref{schottPi} as a total time derivative plus
radiation-reaction terms. The time derivative will be naturally
interpreted as that of the position of the center of mass $G_i$ of the
matter system (multiplied by its mass) which, for gravitating systems,
is nothing but the mass dipole moment $I_i$, so that we shall recover
Eq.~\eqref{Ii} in the 1PN approximation. Multiplying the continuity
equation~\eqref{cont} by $x_i$ and integrating we get
\begin{equation}
\frac{\ud}{\ud t}\left[\int\ud^3\mathbf{x}\,x_i\left( \sigma -
  \frac{1}{c^2}\sigma_{jj} \right)\right] = \int\ud^3\mathbf{x}\left[
  \sigma_i - \frac{1}{c^2} \,x_i \,\sigma \partial_t\mathcal{V}\right]
+ \overline{\mathcal{O}}\left(\frac{1}{c^{8}}\right) \,.
\end{equation}
Thanks to that relation, we can transform the expression~\eqref{Pi} as
\begin{align}\label{Pi2}
P_i &= \frac{\ud}{\ud t}\left[\int\ud^3\mathbf{x}\,x_i\left( \sigma +
  \frac{1}{c^2}\left[\frac{\sigma U}{2} - \sigma_{jj}
    \right]\right)\right] \nonumber\\& \qquad\qquad + \frac{1}{c^2}
\int\ud^3\mathbf{x}\left[ x_i\,\sigma\,\partial_t V^\text{reac} + 4
  \Bigl(\sigma_i V^\text{reac} - \sigma V^\text{reac}_i\Bigr)\right] +
\delta P^\text{reac}_i +
\overline{\mathcal{O}}\left(\frac{1}{c^{8}}\right)\,.
\end{align}
Next, inserting the values~\eqref{Vmu} of the radiation-reaction
potentials $V^\text{reac}_\mu$ and combining the result with the
expression~\eqref{schottPi}, we observe important simplifications,
modulo total time derivatives. Notably, the contributions from the
current quadrupole moment, as well as those from the extra
integrals~\eqref{KLMi}, can all be absorbed into a total time
derivative which will constitute an analogue of the Schott terms in
the flux-balance equation for the center-of-mass position. Finally, we
find that Eqs.~\eqref{Pi}--\eqref{schottPi} can equivalently be
rewritten as
\begin{equation}\label{dGidt}
\frac{\ud G_i}{\ud t} = P_i - \frac{2G}{21 c^7}
\!\!\stackrel{(3)}{I}_{\!\!ijk}\stackrel{(3)}{I}_{\!\!jk} +
\mathcal{O}\left(\frac{1}{c^9}\right)\,,
\end{equation}
where $G_i$ is given by
\begin{equation}\label{Gi}
G_i = \int\ud^3\mathbf{x}\,x_i\left( \sigma +
\frac{1}{c^2}\left[\frac{\sigma U}{2} - \sigma_{jj} \right]\right) +
\delta G_i^\text{reac} +
\overline{\mathcal{O}}\left(\frac{1}{c^{8}}\right)\,.
\end{equation}
We recover, as expected, the expression of the mass-type dipole
moment~\eqref{Ii}.  The radiation-reaction terms $\delta
G_i^\text{reac}$ are given by
\begin{align}\label{schottIi}
\delta G^\text{reac}_i &=
\frac{2G}{5c^5}\left(I_{j}\!\!\stackrel{(3)}{I}_{\!\!ij} -
2\!\!\stackrel{(1)}{I}_{\!\!j} \stackrel{(2)}{I}_{\!\!ij}\right)
\nonumber\\ & + \frac{G}{c^7}\biggl[ - \frac{1}{63} \!\!
  \stackrel{(5)}{I}_{\!\!ijk} \!\!I_{jk} + \frac{2}{63}
  \!\!\stackrel{(4)}{I}_{\!\!ijk}\stackrel{(1)}{I}_{\!\!jk} -
  \frac{1}{21}
  \!\!\stackrel{(3)}{I}_{\!\!ijk}\stackrel{(2)}{I}_{\!\!jk} +
  \frac{19}{315}
  \!\!\stackrel{(2)}{I}_{\!\!ijk}\stackrel{(3)}{I}_{\!\!jk} -
  \frac{2}{45}
  \!\!\stackrel{(1)}{I}_{\!\!ijk}\stackrel{(4)}{I}_{\!\!jk}
  + \frac{1}{35}
  I_{ijk}\!\!\stackrel{(5)}{I}_{\!\!jk} \nonumber\\ &\qquad -
  \frac{8}{45}\varepsilon_{ijk}
  \!\!\stackrel{(3)}{J}_{\!\!jl}\stackrel{(1)}{I}_{\!\!kl} +
  \frac{8}{15}\varepsilon_{ijk}
  \!\!\stackrel{(2)}{J}_{\!\!jl}\stackrel{(2)}{I}_{\!\!kl} -
  \frac{16}{45}\varepsilon_{ijk}
  \!\!\stackrel{(1)}{J}_{\!\!jl}\stackrel{(3)}{I}_{\!\!kl} +
  \frac{8}{45}\varepsilon_{ijk} J_{jl} \!\!
  \stackrel{(4)}{I}_{\!\!kl} \nonumber\\ &\qquad + \frac{1}{25}
  N_{j}\!\!\stackrel{(5)}{I}_{\!\!ij} + \frac{8}{15}
  J_{j}\!\!\stackrel{(3)}{J}_{\!\!ij}\biggr]\,.
\end{align}
As usual, the latter terms~\eqref{schottIi}, although of the same PN
order as the flux terms in the RHS of~\eqref{dGidt}, will in fact be
very small in the adiabatic approximation, \textit{i.e.}, when
considered in average over a typical oscillation period of the
system. Finally, as we discussed in the introduction, the
result~\eqref{dGidt} forms an integral part of the description of the
secular evolution of an isolated system due to gravitational
radiation.

\subsection{Energy and angular momentum} 
\label{sec:EJ}
 
For completeness, we also present the balance equations obeyed by the
energy and angular momentum, following~\cite{B97}. With the precision
of the metric~\eqref{metric}, we are able to write the balance
equations at the subleading 1PN order for conservative effects and
subleading 3.5PN order for radiation-reaction effects. These 1PN
relative equations involve the well known 1PN fluxes in the RHS,
\begin{subequations}\label{dEJdt}
\begin{align}
\frac{\ud E}{\ud t} &= - \frac{G}{c^5} \left( \frac{1}{5}
\!\!\stackrel{(3)}{I}_{\!\!ij}\stackrel{(3)}{I}_{\!\!ij} +
\frac{1}{c^2}\left[\frac{1}{189}
  \!\!\stackrel{(4)}{I}_{\!\!ijk}\stackrel{(4)}{I}_{\!\!ijk} +
  \frac{16}{45} \!\stackrel{(3)}{J}_{\!\!ij}\stackrel{(3)}{J}_{\!\!ij}
  \right] \right) + \mathcal{O}\left( \frac{1}{c^8}\right)
\,, \label{dEdt}\\ \frac{\ud J_i}{\ud t} &= - \frac{G}{c^5}
\varepsilon_{ijk} \left( \frac{2}{5} \!\!
\stackrel{(2)}{I}_{\!\!jl}\stackrel{(3)}{I}_{\!\!kl} +
\frac{1}{c^2}\left[ \frac{1}{63} \!\!
  \stackrel{(3)}{I}_{\!\!jlm}\stackrel{(4)}{I}_{\!\!klm} +
  \frac{32}{45} \!\stackrel{(2)}{J}_{\!\!jl}
  \stackrel{(3)}{J}_{\!\!kl} \right] \right) + \mathcal{O}\left(
\frac{1}{c^8}\right) \,, \label{dJdt}
\end{align}
\end{subequations}
where the 1PN mass quadrupole moment $I_{ij}$ is given
by~\eqref{Iij}. The quantities in the LHS, with consistent accuracy,
read
\begin{subequations}\label{exprEJ}
\begin{align}
E &= \int \ud^3 \mathbf{x} \left( \sigma c^2 + \frac{1}{2} \sigma U -
\sigma_{ii} + \frac{1}{c^2} \biggl[ -4\sigma W_{ii} + 2\sigma_i U_i +
  \frac{1}{2} \sigma \,\partial^2_t X - \frac{1}{4} \partial_t \sigma
  \,\partial_t X \biggr] \right) \nonumber\\ &+ \delta E^\text{reac} +
\overline{\mathcal{O}}\left(\frac{1}{c^{8}}\right)\,, \label{exprE}
\\ J_i &= \varepsilon_{ijk} \int \ud^3 \mathbf{x} \,x_j \left(
\sigma_k + \frac{1}{c^2} \left[ 4 \sigma_k U - 4\sigma U_k -
  \frac{1}{2} \sigma \,\partial_k \partial_t X \right] \right) +
\delta J_i^\text{reac} +
\overline{\mathcal{O}}\left(\frac{1}{c^{8}}\right)\,, \label{exprJ}
\end{align}
\end{subequations}
where $\delta E^\text{reac}$ and $\delta J_i^\text{reac}$ represent
some 2.5PN and 3.5PN contributions coming from total time derivatives
appearing in the RHS, and which will not be needed here. Notice that,
because of the scaling of the rest mass contribution $\sigma c^2$ in
the energy, the relative precision of the expression~\eqref{exprE} is
actually 2PN, \textit{i.e.}, the total mass $M=E/c^2$ is given at the
2PN order. This is why we had to push the accuracy of the spatial part
$g_{ij}$ of the metric~\eqref{metric} up to 2PN order and include the
terms involving the potential $W_{ij}$ defined by Eq.~\eqref{Wij}.

Finally, we recall from~\cite{B97} that we can even include the tail
effect at the 4PN order, since the radiation-reaction potential
$V^\text{reac}$ contains a tail contribution at that order. The 4PN
energy balance equation, for instance, becomes
\begin{align}\label{balancetail}
\frac{\ud E}{\ud t} =& - \frac{G}{5c^5} \left(
\stackrel{(3)}{I}_{\!\!ij}(t) + \frac{G M}{c^3} \int^{+\infty}_0
\ud\tau \ln \tau \left[ \stackrel{(5)}{I}_{\!\!ij} \!\!(t-\tau) +
  \!\!\stackrel{(5)}{I}_{\!\!ij} \!\!(t+\tau) \right]\right)^2
\nonumber \\ & - \frac{G}{c^7} \left[ \frac{1}{189}
  \Bigl(\stackrel{(4)}{I}_{\!\!ijk}\Bigr)^2 + \frac{16}{45}
  \Bigl(\stackrel{(3)}{J}_{\!\!ij}\Bigr)^2 \right] +
\mathcal{O}\left({1\over c^9}\right) \,.
\end{align} 
Note that the tail term in the RHS is given here as an
``antisymmetric'' integral, which corresponds in fact to the
dissipative part of the full tail effect.\footnote{Any constant
  inserted into the logarithmic kernel of that antisymmetric integral
  cancels out.} The energy in the LHS of Eq.~\eqref{balancetail} also
acquires some 4PN tail induced contributions but in the form of some
``symmetric'' integrals, which have been investigated, \textit{e.g.},
in Sec.~IV of~\cite{BBFM17}.
 
\section{Direct computations of the gravitational-wave fluxes}
\label{sec:flux}

In this section, we perform a completely different type of
calculation.  Namely, we compute directly the fluxes appearing in the
RHS of the balance equations from the matter source at future null
infinity $\scri^+$. This will permit to check our new balance
equation~\eqref{dGidt} for the position of the center of mass. In
fact, we shall obtain the full multipole moment expansion for the
flux, going beyond the leading order. We shall also recover the known
multipolar expansions for the fluxes associated with the other
invariants. Our calculation, however, will be restricted to quadratic
order in a post-Minkowskian expansion ($G^2$).
 
\subsection{Integral conservation identities}
\label{sec:cons}

We start with the gauge-fixed Einstein field equations in harmonic
coordinates,
\begin{equation}\label{GFEE}
\Box h^{\mu\nu} = \frac{16\pi G}{c^4}\tau^{\mu\nu}\,,
\end{equation}
where $\Box$ is the flat d'Alembertian operator. In addition, the
field variable $h^{\mu\nu} = \sqrt{-g} g^{\mu\nu} - \eta^{\mu\nu}$
(\textit{i.e.} the ``gothic'' metric deviation from Minkowski's
metric) must satisfy the harmonic gauge condition $\partial_\nu
h^{\mu\nu}=0$. The pseudo stress-energy tensor in the RHS,
\begin{equation}\label{taumunu}
\tau^{\mu\nu} = \vert g\vert T^{\mu\nu} + \frac{c^4}{16\pi G}
\Lambda^{\mu\nu}\,,
\end{equation}
is the sum of the matter contribution with compact support and the
non-linear gravitational source term $\Lambda^{\mu\nu}$, which is at
least quadratic in $h^{\mu\nu}$ and its space-time derivatives.

To derive the flux-balance equations, we integrate the conservation
law $\partial_\nu \tau^{\mu\nu}=0$ (or similar relations following
from it) over a three-dimensional volume $\mathrsfs{V}$ enclosing the
compact-support source and bounded by some two-dimensional surface
$\mathrsfs{S}$. Since we look for the balance equations describing the
evolution, due to gravitational-wave emission, of otherwise constant
quantities, the volume $\mathrsfs{V}$ is chosen to tend asymptotically
toward $\scri^+$. It is thus natural to perform a change of
coordinates $(t,\mathbf{x})\longrightarrow(u,\mathbf{x})$ where $u$
denotes an outgoing null coordinate, satisfying
$g^{\mu\nu}\partial_\mu u\,\partial_\nu u=0$. For simplicity, we take
it to be of the form $u = t - r_*(\mathbf{x})/c$, where the
``tortoise'' coordinate $r_*$ depends on position $\mathbf{x}$ but not
on time $t$. At leading order, the tortoise coordinate contains the
well known logarithmic deviation of light cones in harmonic
coordinates, $r_*=r+\frac{2G M}{c^2}\ln r+\mathcal{O}(r^0)$.  Posing
$n_*^i=\partial_i r_*$ we can rewrite the conservation law of the
pseudo tensor $\tau^{\mu\nu}$ in the coordinate system
$(u,\mathbf{x})$ as\footnote{The nullity condition of the
  $u$-coordinate, $g^{\mu\nu}\partial_\mu u\,\partial_\nu u=0$,
  translates into a condition on the Euclidean norm of the vector
  $n_*^i$, namely $\bm{n}_*^2 = 1 - h^{00} + 2 h^{0i} n_*^i - h^{ij}
  n_*^i n_*^j$. }
\begin{equation}\label{dtaumunu}
\frac{\partial}{c\partial u}\Bigl[ \tau^{\mu 0}(\mathbf{x},u+r_*/c) - n_*^i
\tau^{\mu i}(\mathbf{x},u+r_*/c)\Bigr] + \partial_i\Bigl[ \tau^{\mu
  i}(\mathbf{x},u+r_*/c) \Bigr] = 0 \,.
\end{equation}

In this approach, all volume integrals are defined over the finite volume
$\mathrsfs{V}$, while the surface integrals produced when applying Gauss's law
and used to construct the fluxes are computed over the two-dimensional
boundary $\mathrsfs{S}=\partial\mathrsfs{V}$ of that volume. In the end, we
shall formally let the volume grow to infinity and the surface $\mathrsfs{S}$
tend toward $\scri^+$, in the limit $r\to +\infty$ at $u$=const. It turns out
that all the fluxes are convergent in this limit modulo total time
derivatives. Following this procedure, it is straightforward to obtain the
following flux-balance equations\footnote{See standard textbooks such
  as~\cite{LL} and~\cite{PoissonWill} for the derivation of the flux-balance
  equations. Note however that in the book~\cite{PoissonWill} the integration
  is performed over a volume at constant time $t$ instead of constant outgoing
  null coordinate $u$, so that the computed quantities tend towards the ADM
  rather than Bondi values when the surface $\mathrsfs{S}$ goes to infinity.
  Note also that the flux integral for the center of mass we find in
  Eq.~\eqref{fluxCM} differs from Eq.~(6.35) in Ref.~\cite{PoissonWill}.}
\begin{subequations}\label{flux-bal}
\begin{align}
\frac{\ud E}{\ud u} &= - c \oint_\mathrsfs{S} \ud
S_i\,\tau_\text{GW}^{0i}(\mathbf{x},u+r_*/c) \,,\\ \frac{\ud P^i}{\ud
  u} &= - \oint_\mathrsfs{S} \ud
S_j\,\tau_\text{GW}^{ij}(\mathbf{x},u+r_*/c) \,,\\ \frac{\ud J_i}{\ud
  u} &= - \varepsilon_{ijk} \oint_\mathrsfs{S} \ud S_l\,x^j
\,\tau_\text{GW}^{kl}(\mathbf{x},u+r_*/c)\,,\\ \frac{\ud G_i}{\ud u}
&= P_i - \frac{1}{c} \oint_\mathrsfs{S} \ud S_j \left(
x^i\,\tau_\text{GW}^{0j} - r_*
\,\tau_\text{GW}^{ij}\right)(\mathbf{x},u+r_*/c)\,,\label{fluxCM}
\end{align}
\end{subequations}
where we have denoted quite naturally $\tau_\text{GW}^{\mu\nu} =
\frac{c^4}{16\pi G}\Lambda^{\mu\nu}$. The quantities on the LHS are
the Bondi-like energy, linear momentum, angular momentum (or
current-type dipole moment) and center-of-mass position (or mass-type
dipole moment):
\begin{subequations}\label{Bondi}
\begin{align} 
E &= \int_\mathrsfs{V} \ud^3\mathbf{x} \Bigl[\tau^{00} -
  n_*^i\,\tau^{0i} \Bigr](\mathbf{x},u+r_*/c) \,,\\ P^i &= \frac{1}{c}
\int_\mathrsfs{V} \ud^3\mathbf{x} \Bigl[\tau^{i0} - n_*^j\,\tau^{ij}
  \Bigr](\mathbf{x},u+r_*/c) \,,\\ J_i &= \frac{1}{c}
\,\varepsilon_{ijk} \int_\mathrsfs{V} \ud^3\mathbf{x} \,x^j
\Bigl[\tau^{k 0} - n_*^l\,\tau^{kl} \Bigr](\mathbf{x},u+r_*/c)
\,,\label{defJi}\\ G_i &= \frac{1}{c^2} \,\int_\mathrsfs{V} \ud^3\mathbf{x} \Bigl[
  x^i\bigl(\tau^{00} - n_*^j\,\tau^{0j}\bigr) - r_*\bigl(\tau^{i0} -
  n_*^j\,\tau^{ij}\bigr) \Bigr](\mathbf{x},u+r_*/c)\,.\label{defGi}
\end{align}
\end{subequations}

\subsection{Multipolar expansion of the fluxes}
\label{sec:mult}

Next, we compute the fluxes in the RHS of Eqs.~\eqref{flux-bal}. The
usual way to proceed consists of expanding the Einstein field
equations when $r\to+\infty$ near $\scri^+$.\footnote{See most
  references in the field, from historical works,
  \textit{e.g.},~\cite{BBM62}~\cite{Sachs62}~\cite{Th80}, till recent
  contributions, \textit{e.g.},~\cite{BP18}.} This calculation is then
``exact'', as it requires only the leading and subleading expansion
coefficients $1/r^2$ and $1/r^3$ of the GW stress-energy pseudo tensor
in an appropriate radiative-type coordinate system. However, the
fluxes are then given in terms of the radiative multipole moments (say
$U_L$ and $V_L$), which represent a mere parametrization of the
asymptotic waveform, disconnected from the matter source at this
stage.

Here we adopt a different approach. Namely, we restrict the
computation of the fluxes to dominant order in a post-Minkowskian
expansion ($G\to 0$), which amounts to considering only the quadratic
non-linearities into the GW stress-energy pseudo tensor. The advantage
is that the fluxes are now given in terms of some
\textit{source-rooted} multipole moments, say $M_L$ and $S_L$, which
are well controlled since they admit closed form expressions as
integrals over the source, \textit{i.e.}, over the components of the
pseudo tensor~\cite{BlanchetLR}. In the end, the multipole moment
expansions we get should have the same structure as the ``exact''
fluxes written in terms of the radiative moments $U_L$ and
$V_L$. Higher order effects such as tails can also be incorporated,
with some more work, going to next order in $G$.

Let us notice that, to dominant order in $G$, we do not need to
consider the logarithmic deviation of retarded null cones in harmonic
coordinates, so that we can approximate $r_*$ by $r$ and $n_*^i$ by
$n^i$. The GW pseudo tensor then reads at quadratic order
\begin{equation}\label{tauGWquad}
\tau^{\mu\nu}_\text{GW} = \frac{G c^4}{16\pi}
\,\Lambda_{2}^{\mu\nu}\bigl[h_{1},h_{1}\bigr] + \mathcal{O}\left(G^2\right)\,,
\end{equation}
where $\Lambda_{2}^{\mu\nu}$ is the quadratic piece of the non-linear
source term in Eq.~\eqref{taumunu}, with schematic form
$\Lambda_{2}[h,h] \sim h\partial^2 h+\partial h\partial h$. It is
obtained by inserting the expression of the \textit{linearized} metric
in harmonic coordinates, $G h_{1}^{\mu\nu}$, given outside the matter
source in the form of a full multipole moment expansion, through which
the multipole moments of the source are precisely defined. This
linearized multipolar solution of the vacuum field equations
reads~\cite{Th80}
\begin{subequations}\label{h1munu}
\begin{align}
h^{00}_{1} &= -\frac{4}{c^2}\sum_{\ell=0}^{+\infty}
\frac{(-)^\ell}{\ell !} \partial_L \left( \frac{1}{r} M_L
(u)\right)\,, \\
%%%%%%%%%%%%%%%%%%%%%%%%%%%%%%%%%%%%%%%%%%%
h^{0i}_{1} &= \frac{4}{c^3}\sum_{\ell=1}^{+\infty}
\frac{(-)^\ell}{\ell !}  \left\{ \partial_{L-1} \left( \frac{1}{r}
\!\stackrel{(1)}{M}_{\!iL-1}\!\!(u)\right) + \frac{\ell}{\ell+1}
\varepsilon_{ijk} \partial_{jL-1} \left( \frac{1}{r} S_{kL-1}
(u)\right)\right\}\,, \\
%%%%%%%%%%%%%%%%%%%%%%%%%%%%%%%%%%%%%%%%%%%%%%
h^{ij}_{1} &= -\frac{4}{c^4}\sum_{\ell=2}^{+\infty}
\frac{(-)^\ell}{\ell !}  \left\{ \partial_{L-2} \left( \frac{1}{r}
\!\stackrel{(2)}{M}_{\!ijL-2}\!\!(u)\right) + \frac{2\ell}{\ell+1}
\partial_{kL-2} \left( \frac{1}{r} \varepsilon_{kl(i}
\!\stackrel{(1)}{S}_{\!j)L-2}\!\!(u)\right)\right\}\,.
\end{align}
\end{subequations}
Here, $M_L$ and $S_L$ denote the so-called ``canonical'' mass-type and
current-type multipole moments of the source,\footnote{The canonical
  multipole moments $M_L$ and $S_L$ differ from the source moments
  $I_L$ and $J_L$ by small 2.5PN corrections (see~\cite{BlanchetLR}
  for details):
$$M_L = I_L + \mathcal{O}\left(\frac{1}{c^5}\right)\,,\qquad S_L = J_L
  + \mathcal{O}\left(\frac{1}{c^5}\right)\,.$$
} taken at retarded time $u=t-r$, which agrees in the present
approximation with a true null coordinate. Among those moments, the
mass monopole $M$ and current dipole $S_i$ are constant, while the
mass dipole $M_i$ varies linearly with time, which means that
$\Pi_i\equiv M_i^{(1)}$ is actually constant. The latter monopole and
dipoles thus represent ADM quantities. Now, we need the leading and
subleading order terms, proportional to $1/r$ and $1/r^2$
respectively, in the expansion of $h_{1}^{\mu\nu}$ when $r\to+\infty$
(with $u=\text{const}$),
\begin{equation}\label{h1exp}
h_{1}^{\mu\nu} = \frac{1}{r} \,z_{1}^{\mu\nu}(\bm{n},u) +
\frac{1}{r^2} \, y_{1}^{\mu\nu}(\bm{n},u) +
\mathcal{O}\left(\frac{1}{r^3}\right)\,.
\end{equation}
The coefficients $z_{1}^{\mu\nu}$ and $y_{1}^{\mu\nu}$, which depend
only on the unit direction $\bm{n}$ and the retarded time $u$, are
explicitly given by
\begin{subequations}\label{zmunu}
\begin{align}
z_{1}^{00} &= - 4 \sum_{\ell=0}^{+\infty} \frac{n_L}{\ell!
  c^{\ell+2}} \!  \stackrel{(\ell)}{M}_{\!L}\,,\\ z_{1}^{0i} &= - 4
\sum_{\ell=1}^{+\infty} \frac{n_{L-1}}{\ell! c^{\ell+2}} \!
\stackrel{(\ell)}{M}_{\!iL-1} + 4 \sum_{\ell=1}^{+\infty}
\frac{\ell}{(\ell+1)! c^{\ell+3}}\,\varepsilon_{ijk}\,n_{jL-1} \!
\stackrel{(\ell)}{S}_{\!kL-1}\,,\\ z_{1}^{ij} &= - 4
\sum_{\ell=2}^{+\infty} \frac{n_{L-2}}{\ell! c^{\ell+2}} \!
\stackrel{(\ell)}{M}_{\!ijL-2} + 8 \sum_{\ell=2}^{+\infty}
\frac{\ell}{(\ell+1)! c^{\ell+3}}\,n_{kL-2}\,\varepsilon_{kl(i} \!
\stackrel{(\ell)}{S}_{\!j)lL-2}\,,
\end{align}
\end{subequations}
and
\begin{subequations}\label{ymunu}
\begin{align}
y_{1}^{00} &= - 2 \sum_{\ell=1}^{+\infty} \frac{\ell(\ell+1)}{\ell!
  c^{\ell+1}} \,n_L
\!\!\!\stackrel{(\ell-1)}{M}_{\!\!\!L}\,,\\ y_{1}^{0i} &= - 2
\sum_{\ell=2}^{+\infty} \frac{(\ell-1)\ell}{\ell!  c^{\ell+1}}
\,n_{L-1} \!\!\!\stackrel{(\ell-1)}{M}_{\!\!\!iL-1} + 2
\sum_{\ell=1}^{+\infty} \frac{\ell^2(\ell+1)}{(\ell+1)!
  c^{\ell+2}}\,\varepsilon_{ijk}\,n_{jL-1}
\!\!\!\stackrel{(\ell-1)}{S}_{\!\!\!\!kL-1}\,,\\ y_{1}^{ij} &= - 2
\sum_{\ell=3}^{+\infty} \frac{(\ell-2)(\ell-1)}{\ell!  c^{\ell+1}}
\,n_{L-2} \!\!\!\stackrel{(\ell-1)}{M}_{\!\!\!ijL-2} + 4
\sum_{\ell=2}^{+\infty} \frac{(\ell-1)\ell^2}{(\ell+1)!
  c^{\ell+2}}\,n_{kL-2}\,\varepsilon_{kl(i}
\!\!\!\stackrel{(\ell-1)}{S}_{\!\!\!\!j)lL-2}\,.
\end{align}
\end{subequations}
We plug these expressions into the quadratic source
$\Lambda_{2}^{\mu\nu}$ to control the leading and subleading terms,
behaving as $1/r^2$ and $1/r^3$ respectively, of its asymptotic
expansion in powers of $1/r$ at null infinity
\begin{equation}\label{Lambda2exp}
\Lambda_{2}^{\mu\nu} = \frac{1}{r^2} \,Q_{2}^{\mu\nu}(\bm{n},u) +
\frac{1}{r^3} \,R_{2}^{\mu\nu}(\bm{n},u) +
\mathcal{O}\left(\frac{1}{r^4}\right)\,.
\end{equation}
Inserting then these expansions into the fluxes~\eqref{flux-bal} and
integrating over a sphere at infinity (hence $\ud S_i = \ud \Omega
\,n_i\,r^2$ with $\ud \Omega$ representing the element of solid angle)
yields
\begin{subequations}\label{flux-bal2}
\begin{align}
\frac{\ud E}{\ud u} &= - \frac{G c^5}{16\pi} \oint_\mathrsfs{S} \ud
\Omega \,n_i\,Q_{2}^{0i} + \mathcal{O}\left(\frac{1}{r}\right) +
\mathcal{O}\left(G^2\right) \,,\\ \frac{\ud P^i}{\ud u} &= - \frac{G
  c^4}{16\pi} \oint_\mathrsfs{S} \ud \Omega \,n_j\,Q_{2}^{ij} +
\mathcal{O}\left(\frac{1}{r}\right) +
\mathcal{O}\left(G^2\right)\,,\\ \frac{\ud J_i}{\ud u} &= - \frac{G
  c^4}{16\pi} \,\varepsilon_{ijk} \oint_\mathrsfs{S} \ud \Omega \,n_j
n_l \Bigl[r\,Q_{2}^{kl} + R_{2}^{kl}\Bigr] +
\mathcal{O}\left(\frac{1}{r}\right) +
\mathcal{O}\left(G^2\right)\,,\label{flux-bal2c}\\ \frac{\ud G_i}{\ud
  u} &= P_i - \frac{G c^3}{16\pi} \,\oint_\mathrsfs{S} \ud \Omega \,n_j
\Bigl[ r \left( n_i\,Q_{2}^{0j} - Q_{2}^{ij} \right) + \left(
  n_i\,R_{2}^{0j} - R_{2}^{ij} \right)\Bigr] +
\mathcal{O}\left(\frac{1}{r}\right) +
\mathcal{O}\left(G^2\right)\,.\label{flux-bal2d}
\end{align}
\end{subequations}
Importantly, for the fluxes of angular momentum $J_i$ and center-of-mass
position $G_i$, the leading term formally behaves like $r$, implying that the
fluxes \textit{a priori} diverge at infinity. However, we shall find that
those divergent contributions vanish after angular integration, modulo total
time derivatives which reduce to zero in the center-of-mass frame and can be
removed by means of an appropriate redefinition of $J_i$ and $G_i$ (see the
footnote~\ref{footnote}). This fact is well known in the case of the angular
momentum and we will verify it explicitly in the case of the center-of-mass
position.

To proceed further, we denote by $k^\mu=(1,\bm{n})$ a Minkowskian null
vector [with thus $k_\nu=(-1,\bm{n})$], by $n^\mu=(0,\bm{n})$ the
corresponding purely spatial vector, and by $\delta_\nu=(0,\delta_i)$
an operator purely acting on angles and whose space part is defined as
$\delta_i= r\,\partial_i n_j\frac{\partial}{\partial n^j} =
\perp_{ij}\frac{\partial}{\partial n^j}$, where
$\perp_{ij}=\delta_{ij}-n_i n_j$ is the orthogonal projector onto the
plane perpendicular to $n_i$. The following relations then hold:
$k^\mu k_\mu=0$, $n^\mu k_\mu=1$ and $k^\mu \delta_\mu = n^\mu
\delta_\mu =0$.  Moreover, the tensor $\delta_\mu k_\nu$ is actually
symmetric, \textit{i.e.}, $\delta_\mu k_\nu=\delta_\nu k_\mu$, since
its $ij$ components are given by $\perp_{ij}$ whereas its other
components are zero. With those notations, the harmonic-gauge
condition at the linearized order, $\partial_\nu h_1^{\mu\nu}=0$,
implies the constraints
\begin{subequations}\label{harm1}
\begin{align}
k_\nu\!\!\stackrel{(1)}{z}\!\!{}_{1}^{\mu\nu} &=
0\,,\\ k_\nu\!\!\stackrel{(1)}{y}\!\!{}_{1}^{\mu\nu} &= \bigl(
\delta_\nu - n_\nu \bigr) z_{1}^{\mu\nu}\,.
\end{align}
\end{subequations}
After inserting the expressions~\eqref{zmunu}--\eqref{ymunu} into the
quadratic source $\Lambda_{2}$ of the Einstein field equations, we
readily obtain the leading order coefficient of $1/r^2$ when
$r\to+\infty$ [see Eq.~\eqref{Lambda2exp}] as
\begin{equation}\label{Q2}
Q_{2}^{\mu\nu} = - 4 \frac{k^\rho \Pi_\rho}{c^5}
\stackrel{(2)}{z}{}_{\!\!\!\!1}^{\!\mu\nu} + \frac{k^\mu k^\nu}{c^2}
\Bigl(\frac{1}{2} \stackrel{(1)}{z}{}_{\!\!\!\!1}^{\!\rho\sigma}
\!\stackrel{(1)}{z}{}_{\!\!\!\!1\rho\sigma} - \frac{1}{4}
\stackrel{(1)}{z}{}_{\!\!\!\!1\rho}^{\!\rho}
\stackrel{(1)}{z}{}_{\!\!\!\!1\sigma}^{\!\sigma}\Bigr)\,,
\end{equation}
where $\Pi_\rho=(M c,\Pi_i)$ with $\Pi_i=M_i^{(1)}$ denotes the
constant (ADM) linear four-momentum. The quantity in parenthesis in
the second term is proportional to the gravitational-wave energy flux
(at quadratic order). The next order piece, $\propto 1/r^3$, in the
quadratic source $\Lambda_2$ is more involved:
\begin{align}\label{R2}
R_{2}^{\mu\nu} =& - 4 \frac{k^\rho \Pi_\rho}{c^5}
\stackrel{(2)}{y}{}_{\!\!\!\!1}^{\!\mu\nu} - 8 \frac{n^\rho
  \Pi_\rho}{c^5} \stackrel{(1)}{z}{}_{\!\!\!\!1}^{\!\mu\nu} + 8
\Pi^\rho \delta_\rho \stackrel{(1)}{z}{}_{\!\!\!\!1}^{\!\mu\nu} + 4
\eta^{\mu\nu} n_\rho \Pi_\sigma
\stackrel{(1)}{z}{}_{\!\!\!\!1}^{\!\rho\sigma} \\ &+ 8
\stackrel{(1)}{z}{}_{\!\!\!\!1}^{\!\rho(\mu}\left(\Pi^{\nu)}n_\rho-n^{\nu)}\Pi_\rho\right)
+ \delta_\rho k_\sigma z^{\rho\sigma}
\stackrel{(1)}{z}{}_{\!\!\!\!1}^{\!\mu\nu} - k_\rho k_\sigma
y^{\rho\sigma} \stackrel{(2)}{z}{}_{\!\!\!\!1}^{\!\mu\nu} +
\eta^{\mu\nu} \delta_\rho k_\sigma
\stackrel{(1)}{z}{}_{\!\!\!\!1}^{\!\rho\lambda}
z^\sigma_{1\lambda}\nonumber\\ & - \eta^{\mu\nu}\left( \frac{1}{2}
\stackrel{(1)}{z}{}_{\!\!\!\!1}^{\!\rho\sigma} \,z_{1\rho\sigma} -
\frac{1}{4} \stackrel{(1)}{z}{}_{\!\!\!\!1\rho}^{\!\rho}
z_{1\sigma}^\sigma \right) + k^{\mu}k^{\nu}\left(
\stackrel{(1)}{z}{}_{\!\!\!\!1}^{\!\rho\sigma}
\!\stackrel{(1)}{y}{}_{\!\!\!\!1\rho\sigma} - \frac{1}{2}
\stackrel{(1)}{z}{}_{\!\!\!\!1\rho}^{\!\rho}
\stackrel{(1)}{y}{}_{\!\!\!\!1\sigma}^{\!\sigma}\right) + \frac{1}{2}
\!\stackrel{(1)}{z}{}_{\!\!\!\!1\rho}^{\!\rho}\,k^{(\mu}
\,\delta^{\nu)} z^{\sigma}_{1\sigma} \nonumber\\ & +
k^{(\mu}n^{\nu)}\left( \stackrel{(1)}{z}{}_{\!\!\!\!1}^{\!\rho\sigma}
z_{1\rho\sigma} - \frac{1}{2}
\stackrel{(1)}{z}{}_{\!\!\!\!1\rho}^{\!\rho}
\,z^{\sigma}_{1\sigma}\right) + 2 \delta_\rho k_\sigma
\stackrel{(1)}{z}{}_{\!\!\!\!1}^{\!\rho(\mu} z_1^{\nu)\sigma} + 2
\stackrel{(1)}{z}{}_{\!\!\!\!1}^{\!\rho(\mu} z^{\nu)}_{1\rho} - 2
\delta^{(\mu}k_\rho \stackrel{(1)}{z}{}_{\!\!\!\!1}^{\!\nu)\sigma}
z^{\rho}_{1\sigma} \nonumber\\ &~ -
\stackrel{(1)}{z}{}_{\!\!\!\!1}^{\!\rho\sigma} k^{(\mu} \delta^{\nu)}
z_{1\rho\sigma} + 2 k^{(\mu}
\stackrel{(1)}{z}{}_{\!\!\!\!1}^{\!\rho\sigma} \delta_\rho
z^{\nu)}_{1\sigma} - 2 k^{(\mu}
\stackrel{(1)}{z}{}_{\!\!\!\!1\rho\sigma} n^{\rho}
z_1^{\nu)\sigma} - 2 \stackrel{(1)}{z}{}_{\!\!\!\!1}^{\!\rho(\mu}
k^{\nu)} \left(\delta_\sigma - n_\sigma\right) z^\sigma_{1\rho}
\,.\nonumber
\end{align}
As a check of those expressions, due to the fact that $\partial_\nu
\Lambda_2^{\mu\nu}=0$, we must have
\begin{subequations}\label{harm2}
\begin{align}
k_\nu\!\!\stackrel{(1)}{Q}\!\!{}_{2}^{\mu\nu} &=
0\,,\\ k_\nu\!\!\stackrel{(1)}{R}\!\!{}_{2}^{\mu\nu} &= \bigl(
\delta_\nu - 2 n_\nu \bigr) Q_{2}^{\mu\nu}\,.
\end{align}
\end{subequations}

The final steps consist of replacing ${Q}_{2}^{\mu\nu}$ and
${R}_{2}^{\mu\nu}$ by their expressions~\eqref{Q2}--\eqref{R2} in
Eqs.~\eqref{flux-bal2}, to perform the angular integration and let the
surface $\mathrsfs{S}$ tend towards $\scri^+$. For the fluxes of $E$
and $P_i$, the computation is straightforward and not too long,
leading to the well-known multipolar series parametrized by the
(canonical) mass and current multipole moments $M_L$ and $S_L$:
\begin{subequations}\label{FluxesEP}\begin{align}
\frac{\ud E}{\ud u} &= - \sum^{+\infty}_{\ell=2} \frac{G}{c^{2\ell+1}}
\biggl\{ \frac{(\ell+1)(\ell+2)}{(\ell-1)\ell \ell!(2\ell+1)!!}
\!\!\stackrel{(\ell+1)}{M}_{\!\!\!L} \stackrel{(\ell+1)}{M}_{\!\!\!L}
\nonumber\\ & \qquad\qquad\qquad+ \frac{4\ell
  (\ell+2)}{c^2(\ell-1)(\ell+1)!(2\ell+1)!!}
\!\!\stackrel{(\ell+1)}{S}_{\!\!\!\! L}
\stackrel{(\ell+1)}{S}_{\!\!\!\! L}\biggr\} +
\mathcal{O}\left(G^2\right)\,,\label{FluxE}\\ \frac{\ud P_i}{\ud u} &=
- \sum^{+\infty}_{\ell=2} \frac{G}{c^{2\ell+3}} \biggl\{
\frac{2(\ell+2)(\ell+3)}{\ell(\ell+1)!(2\ell+3)!!}
\!\!\stackrel{(\ell+2)}{M}_{\!\!\!iL} \stackrel{(\ell+1)}{M}_{\!\!\!L}
+ \frac{8(\ell+2)}{(\ell-1)(\ell+1)!(2\ell+1)!!}  \,\varepsilon_{ijk}
\!\!\stackrel{(\ell+1)}{M}_{\!\!\!jL-1}
\!\stackrel{(\ell+1)}{S}_{\!\!\!\!kL-1} \nonumber\\ & \qquad\qquad\qquad
 + \frac{8(\ell+3)}{c^2(\ell+1)!(2\ell+3)!!}
\!\!\stackrel{(\ell+2)}{S}_{\!\!\!\!iL}
\!\stackrel{(\ell+1)}{S}_{\!\!\!\!L}\biggr\} +
\mathcal{O}\left(G^2\right)\,.\label{FluxP}
\end{align}\end{subequations}
Here the moments are evaluated at the retarded time $u$, which becomes a
retarded null coordinate with the present approximation, where contributions
to $\tau^{\mu\nu}_\text{GW}$ beyond quadratic order in $G$ are ignored. In the
case of the linear momentum we discard total time derivatives since they can
be transferred to the LHS of the balance equations.

Concerning the fluxes of $J_i$ and $G_i$, the calculations are more involved
because of the more complicated structure of the term $R_{2}$ given
by~\eqref{R2}.\footnote{The explicit computations of the fluxes entering the
  RHS of~\eqref{flux-bal2} leading to Eqs.~\eqref{FluxesEP}--\eqref{FluxesJG}
  have been performed with the software
  \textit{Mathematica}\hspace{0.5ex}{\scriptsize \textregistered}.}
Furthermore, we have to check that the formally divergent parts of the fluxes
when $r\to +\infty$ [terms involving an explicit factor $r$
in~\eqref{flux-bal2c} and~\eqref{flux-bal2d}] actually reduce to total time
derivatives after angular averaging. In fact, we find the very satisfying
result that all the divergent terms cancel out after the angular integration
when the source is at rest, $\Pi_i = 0$.\footnote{When the source is moving
  with respect to the asymptotic rest frame (\textit{i.e.}, $\Pi_i\not= 0$),
  we find some remaining divergent terms, but only in the form of a total time
  derivative. These terms correct the angular momentum $J_i$ and center of
  mass $G_i$ in the LHS by the quantities:
\begin{align*}
\delta J_i^\text{div} &= \frac{4 G}{15 c^6} \,r \!\stackrel{(3)}{S}_{\!ij} \Pi_j\,,\\
\delta G_i^\text{div} &= - \frac{2 G}{5 c^6} \,r \!\stackrel{(3)}{M}_{\! ij}\Pi_j\,.
\end{align*}
We leave to future work the task of investigating how these divergent terms
should combine with other divergences in the definitions of $J_i$ and $G_i$
[Eqs.~\eqref{defJi} and~\eqref{defGi}] to yield a finite
result.\label{footnote}}
Once this
verification has been done, supplemented with an appropriate redefinition of $J_i$
and $G_i$ in the LHS, we can take the limit $r\to+\infty$. We find
\begin{subequations}\label{FluxesJG}\begin{align}
\frac{\ud J_i}{\ud u} &= - \varepsilon_{ijk} \sum^{+\infty}_{\ell=2}
\frac{G}{c^{2\ell+1}} \biggl\{
\frac{(\ell+1)(\ell+2)}{(\ell-1)\ell!(2\ell+1)!!}
\!\stackrel{(\ell)}{M}_{\!jL-1} \!\stackrel{(\ell+1)}{M}_{\!\!\!kL-1}
\nonumber\\ & \qquad\qquad + \frac{4\ell^2
  (\ell+2)}{c^2(\ell-1)(\ell+1)!(2\ell+1)!!}
\!\stackrel{(\ell)}{S}_{\!jL-1}
\!\stackrel{(\ell+1)}{S}_{\!\!\!\!kL-1}\biggr\} +
\mathcal{O}\left(G^2\right)\,,\label{FluxJ}\\ \frac{\ud G_i}{\ud u} &=
P_i - \sum^{+\infty}_{\ell=2} \frac{G}{c^{2\ell+3}}
\biggl\{\frac{2(\ell+2)(\ell+3)}{\ell\,\ell!(2\ell+3)!!}\!
\stackrel{(\ell+1)}{M}_{\!\!\!iL} \stackrel{(\ell+1)}{M}_{\!\!\!L}+
\frac{8(\ell+3)}{c^2\ell!(2\ell+3)!!}\!
\stackrel{(\ell+1)}{S}_{\!\!\!\!iL}
\stackrel{(\ell+1)}{S}_{\!\!\!\!L}\biggr\} +
\mathcal{O}\left(G^2\right)\,.\label{FluxG}
\end{align}\end{subequations}

As we stressed several times, the
formulas~\eqref{FluxesEP}--\eqref{FluxesJG} are approximate, but the
multipole moments $M_L$ and $S_L$ therein can be related to the matter
source in a precise way; in particular they differ from the source
moments $I_L$ and $J_L$ by small 2.5PN corrections. The next-order
level $\mathcal{O}(G^2)$ contains the tail effect, which can be
included if necessary into the definition of the radiative multipole
moments $U_L$ and $V_L$ measured at $\scri^+$. An alternative
derivation of the fluxes, directly in terms of the radiative moments
$U_L$ and $V_L$, should yield the same structure as
in~\eqref{FluxesEP}--\eqref{FluxesJG}, but with the replacements
$M_L^{(\ell)}\rightarrow U_L$ and $S_L^{(\ell)}\rightarrow V_L$. The
formula~\eqref{FluxJ} for the angular momentum (or, rather, its
equivalent in terms of radiative moments) is already known (see
notably~\cite{Th80}).\footnote{This formula has been provided by
  Thorne~\cite{Th80} (who refers also to an unpublished calculation by
  DeWitt) without details. Here, we have tried to be more
  comprehensive by providing in Eq.~\eqref{R2} the explicit expression
  of the crucial $1/r^3$ piece of the GW pseudo tensor. See Ref.~\cite{BP18}
  for a recent investigation of the angular momentum flux.}

As for the formula~\eqref{FluxG} regarding the center of mass position, it
does not seem to have appeared previously in the literature. For $\ell=2$, we
recover the dominant mass-type contribution~\eqref{balanceG}, with the correct
coefficient $2/21$ derived by means of a radiation-reaction calculation in
Sec.~\eqref{sec:balance}. This coefficient has also been found in
Ref.~\cite{KNQ18}, as well as the dominant current-type contribution
in the case $\ell=2$, in agreement with our formula~\eqref{FluxG} (modulo a
time derivative). We shall now present still another confirmation of the
leading effect~\eqref{balanceG} by a radiation-reaction calculation but, this
time, restricted to the case of compact binary systems.

\section{Radiation-reaction force for compact binary systems}
\label{sec:eom}

The radiation-reaction force on compact binary systems at the 3.5PN
order has been investigated in many works. It was computed in a large
class of coordinate systems, though limiting oneself to the frame of
the center of mass, as a consequence of the energy and angular
momentum balance equations~\cite{IW93,IW95}. On the other hand, using
various approaches and specific coordinate systems, it was established
in a general frame from first principles, \textit{i.e.}, not relying
on any balance equation~\cite{JaraS97,PW02,KFS03,NB05,itoh3}. All
these works consistently agree, as they do with the general fluid
formalism described in Sec.~\ref{sec:radreac}.

The two mass components are referred to as $m_1$ and $m_2$
henceforth. We denote by $r_{12}=\vert\bm{y}_1-\bm{y}_2\vert$ the
harmonic-coordinate distance between the two particles
$\{\bm{y}_1,\bm{y}_2\}$, by $\bm{n}_{12}=(\bm{y}_1-\bm{y}_2)/r_{12}$
the corresponding unit direction, by $\bm{v}_1=\ud \bm{y}_1/\ud t$ and
$\bm{a}_1=\ud \bm{v}_1/\ud t$ the coordinate velocity and acceleration
of the particle 1, respectively (\textit{idem} for particle 2). We
shall also occasionally use the notation
$\bm{v}_{12}=\bm{v}_1-\bm{v}_2$ for the relative velocity. The
Euclidean scalar product of vectors is denoted with parentheses,
\textit{e.g.}  $(n_{12}v_1)=\bm{n}_{12}\cdot\bm{ v}_1$. The 3.5PN
acceleration of the particle 1 can then be written, for general orbits
in a general harmonic-coordinate system, as
\begin{equation}\label{acc1}
\bm{a}_1 = \bm{a}_{1}^\mathrm{N} +
\frac{1}{c^2}\bm{a}_{1}^\mathrm{1PN} +
\frac{1}{c^4}\bm{a}_{1}^\mathrm{2PN} +
\frac{1}{c^5}\bm{a}_{1}^\mathrm{2.5PN} +
\frac{1}{c^6}\bm{a}_{1}^\mathrm{3PN} +
\frac{1}{c^7}\bm{a}_{1}^\mathrm{3.5PN} +
\mathcal{O}\left(\frac{1}{c^8}\right)\,.
\end{equation}
The conservative part of the acceleration is actually known up to the
4PN order~\cite{BBFM17}. The dissipative radiation-reaction part of
interest here reads, at 2.5PN and 3.5PN orders,
\begin{subequations}\label{A25PN35PN}
\begin{align}
\bm{a}_{1}^\mathrm{2.5PN} &= \frac{4 \,G^2 m_1 m_2}{5 \,r_{12}^3}
\left( (n_{12} v_{12}) \left[ - 6 \,\frac{G m_1}{r_{12}} +
  \frac{52}{3} \,\frac{G m_2}{r_{12}} + 3 \,v_{12}^2
  \right]\bm{n}_{12} \right.  \nonumber\\ & \qquad\qquad\quad \left. +
\left[ 2 \, \frac{G m_1}{r_{12}} - 8 \, \frac{G m_2}{r_{12}} -
  v_{12}^2 \right]\bm{v}_{12} \right)\,,
\label{a1_5}\\
%%%%%%%%%%%%%%%%%%%%%%%%%%%%%%%%%%%%%%%%%%%%
%%%%%%%%%%%%%%%%%%%%%%%%%%%%%%%%%%%%%%%%%%%%
\bm{a}_{1}^\mathrm{3.5PN} &=
\frac{G^2 m_1 m_2}{r_{12}^3} \, \left\{ \frac{G^2 m_1^2}{r_{12}^2} \, \left[
\left( \, \frac{3992}{105} (n_{12} v_1) - \frac{4328}{105} (n_{12} v_2)
\right)\bm{n}_{12} - \frac{184}{21} \bm{v}_{12} \, \right]
\right.\nonumber\\
%%%%%%%%%%%%%%%%%%%%%%%%%%%%%%%%%%%%%%%%%%%%
%%%%%%%%%%%%%%%%%%%%%%%%%%%%%%%%%%%%%%%%%%%%
& + \frac{G^2 m_1 m_2}{r_{12}^3} \, \left[ \left( \, -
  \frac{13576}{105} (n_{12} v_1) \, + \, \frac{2872}{21} (n_{12} v_2)
  \, \right)\bm{n}_{12} + \frac{6224}{105} \bm{v}_{12} \, \right]
\nonumber\\
%%%%%%%%%%%%%%%%%%%%%%%%%%%%%%%%%%%%%%%%%%%%
%%%%%%%%%%%%%%%%%%%%%%%%%%%%%%%%%%%%%%%%%%%%
& + \frac{G^2 m_2^2}{r_{12}^3} \, \left[ - \frac{3172}{21} (n_{12}
  v_{12})\,\bm{n}_{12} + \frac{6388}{105} \bm{v}_{12} \right]
\nonumber\\
%%%%%%%%%%%%%%%%%%%%%%%%%%%%%%%%%%%%%%%%%%%%
%%%%%%%%%%%%%%%%%%%%%%%%%%%%%%%%%%%%%%%%%%%%
& + \frac{G m_1}{r_{12}} \, \left[ \left( \, 48 (n_{12} v_1)^3 -
  \frac{696}{5} (n_{12} v_1)^2 (n_{12} v_2) + \frac{744}{5} (n_{12}
  v_1) (n_{12} v_2)^2 \right.\right.\nonumber\\ &\quad\qquad\qquad
  \left.\left. - \frac{288}{5} (n_{12} v_2)^3 -\frac{4888}{105}
  (n_{12} v_1) v_1^2 + \frac{5056}{105} (n_{12} v_2) v_1^2
  \right.\right.\nonumber\\ &\quad\qquad\qquad \left.\left. +
  \frac{2056}{21} (n_{12} v_1) (v_1 v_2) - \frac{2224}{21} (n_{12}
  v_2) (v_1 v_2) \right.\right.\nonumber\\ &\quad\qquad\qquad
  \left.\left. - \frac{1028}{21} (n_{12} v_1) v_2^2 + \frac{5812}{105}
  (n_{12} v_2) v_2^2 \right)\bm{n}_{12} \right. \nonumber\\
%%%%%%%%%%%%%%%%%
&\quad\qquad \left. + \left( \frac{52}{15} (n_{12} v_1)^2 -
  \frac{56}{15} (n_{12} v_1) (n_{12} v_2) - \frac{44}{15} (n_{12}
  v_2)^2 - \frac{132}{35} v_1^2
  \right.\right. \nonumber\\ &\quad\qquad\qquad \left.\left. +
  \frac{152}{35} (v_1 v_2) - \frac{48}{35} v_2^2 \right)\bm{v}_{12}
  \right] \nonumber\\
%%%%%%%%%%%%%%%%%%%%%%%%%%%%%%%%%%%%%%%%%%%%
%%%%%%%%%%%%%%%%%%%%%%%%%%%%%%%%%%%%%%%%%%%%
& + \frac{G m_2}{r_{12}} \, \left[ \left( - \frac{582}{5} (n_{12}
  v_1)^3 + \frac{1746}{5} (n_{12} v_1)^2 (n_{12} v_2) - \frac{1954}{5}
  (n_{12} v_1) (n_{12} v_2)^2
  \right.\right.\nonumber\\ &\quad\qquad\qquad \left.\left. + \, 158
  (n_{12} v_2)^3 +\frac{3568}{105} (n_{12} v_{12}) (v_1 v_1) -
  \frac{2864}{35} (n_{12} v_1) (v_1 v_2)
  \right.\right.\nonumber\\ &\quad\qquad\qquad \left.\left. +
  \frac{10048}{105} (n_{12} v_2) (v_1 v_2) +\frac{1432}{35} (n_{12}
  v_1) v_2^2
%%%%%%%\right.\right.\nonumber\\&\quad\qquad\qquad \left.\left. 
- \frac{5752}{105} (n_{12} v_2) v_2^2 \right)\bm{n}_{12}\right. \nonumber\\
%%%%%%%%%%%%%%%%%%%
&\quad\qquad \left. + \left( \frac{454}{15} (n_{12} v_1)^2 - \frac{372}{5}
(n_{12} v_1) (n_{12} v_2) + \frac{854}{15} (n_{12} v_2)^2 - \frac{152}{21}
v_1^2 \right.\right.\nonumber\\ &\quad\qquad\qquad \left.\left. +
\frac{2864}{105} (v_1 v_2) - \frac{1768}{105} v_2^2 \right)\bm{v}_{12}
\right]\nonumber\\
%%%%%%%%%%%%%%%%%%%%%%%%%%%%%%%%%%%%%%%%%%%
%%%%%%%%%%%%%%%%%%%%%%%%%%%%%%%%%%%%%%%%%%%
& + \left( - 56 (n_{12} v_{12})^5 + 60 (n_{12} v_{1})^3 v_{12}^2 - 180
(n_{12} v_{1})^2 (n_{12} v_{2}) v_{12}^2 + 174 (n_{12} v_{1}) (n_{12}
v_{2})^2 v_{12}^2 \right. \nonumber\\ &\quad\qquad\qquad
\left.\left. - 54 (n_{12} v_{2})^3 v_{12}^2 -\frac{246}{35} (n_{12}
v_{12}) v_1^4 + \frac{1068}{35} (n_{12} v_1) v_1^2 (v_1 v_2)
\right.\right. \nonumber\\ &\quad\qquad\qquad \left.\left. -
\frac{984}{35} (n_{12} v_2) v_1^2 (v_1 v_2) - \frac{1068}{35} (n_{12}
v_1) (v_1 v_2)^2 + \frac{180}{7} (n_{12} v_2) (v_1 v_2)^2
\right.\right. \nonumber\\ &\quad\qquad\qquad \left.\left. -
\frac{534}{35} (n_{12} v_1) v_1^2 v_2^2 + \frac{90}{7} (n_{12} v_2)
v_1^2 v_2^2 + \frac{984}{35} (n_{12} v_1) (v_1 v_2) v_2^2
\right.\right. \nonumber\\ &\quad\qquad\qquad \left.\left. -
\frac{732}{35} (n_{12} v_2) (v_1 v_2) v_2^2 - \frac{204}{35} (n_{12}
v_1) v_2^4 + \frac{24}{7} (n_{12} v_2) v_2^4
\right)\bm{n}_{12}\right. \nonumber\\
%%%%%%%%%%%%%%%%%%%%
& + \left. \left( 60 (n_{12} v_{12})^4 - \frac{348}{5} (n_{12} v_1)^2
v_{12}^2 + \frac{684}{5} (n_{12} v_1) (n_{12} v_2) v_{12}^2 - 66
(n_{12} v_2)^2 v_{12}^2 \right.\right. \nonumber\\ &\quad\qquad\qquad
\left.\left. + \frac{334}{35} v_1^4 - \frac{1336}{35} v_1^2 (v_1 v_2)
+ \frac{1308}{35} (v_1 v_2)^2 + \frac{654}{35} v_1^2 v_2^2
\right.\right. \nonumber\\ &\quad\qquad\qquad \left.\left. -
\frac{1252}{35} (v_1 v_2) v_2^2 + \frac{292}{35} v_2^4
\right)\bm{v}_{12} \right\}\,.\label{a1_7}
\end{align}
\end{subequations}
With this acceleration in hand, it is straightforward to derive the
balance equations. Consistently with the accuracy
of~\eqref{A25PN35PN}, we can work out the 1PN relative equations for
energy and angular momentum [see~\eqref{dEJdt}] as well as the
Newtonian relative ones for linear momentum and center-of-mass
position [see~\eqref{dPidt} and~\eqref{dGidt}]. The test we have made
consists in verifying that, for each of these quantities, there exist
some radiation-reaction terms in the form of total time derivatives
that will contribute to the LHS of the balance equations, with the
expected fluxes in the RHS. In fact, the unique existence of these
terms (in the particular coordinate system we are working with) is
necessary and sufficient to prove the correctness of the balance
equations for the binary systems.

Let us apply the method to the linear momentum $P_i$ and center of
mass position $G_i$, in order to check the coefficient $2/21$ in front
of the new flux term in Eq.~\eqref{dGidt}. We want thus to construct
some total time derivatives so that, as a consequence of the
radiation-reaction force~\eqref{A25PN35PN}, the balance
equations~\eqref{dPidt} and~\eqref{dGidt} are satisfied. We indeed
find well defined expressions for the linear momentum and
center-of-mass position,
\begin{subequations}\label{PIexp}
\begin{align}
\mathbf{P} &= \mathbf{P}^\mathrm{N} +
\frac{1}{c^2}\mathbf{P}^\mathrm{1PN} +
\frac{1}{c^4}\mathbf{P}^\mathrm{2PN} +
\frac{1}{c^5}\mathbf{P}^\mathrm{2.5PN} +
\frac{1}{c^6}\mathbf{P}^\mathrm{3PN} +
\frac{1}{c^7}\mathbf{P}^\mathrm{3.5PN} +
\mathcal{O}\left(\frac{1}{c^8}\right)\,,\\ \mathbf{G} &=
\mathbf{G}^\mathrm{N} + \frac{1}{c^2}\mathbf{G}^\mathrm{1PN} +
\frac{1}{c^4}\mathbf{G}^\mathrm{2PN} +
\frac{1}{c^5}\mathbf{G}^\mathrm{2.5PN} +
\frac{1}{c^6}\mathbf{G}^\mathrm{3PN} +
\frac{1}{c^7}\mathbf{G}^\mathrm{3.5PN} +
\mathcal{O}\left(\frac{1}{c^8}\right)\,,
\end{align}
\end{subequations}
in which the 2.5PN and 3.5PN radiation-reaction terms are uniquely
determined as
\begin{subequations}\label{schottPharm}
\begin{align}
\mathbf{P}^\mathrm{2.5PN} &= \frac{4G^2m_1^2m_2}{5
  r_{12}^2}\left(v_{12}^2 - \frac{2G m_1}{r_{12}}\right)\bm{n}_{12} +
1\leftrightarrow 2\,,\\ \mathbf{P}^\mathrm{3.5PN} &= \biggl
\{\frac{288}{35} \frac{G^4 m_{1}^4 m_{2}}{r_{12}^4} + \frac{332}{35}
\frac{G^4 m_{1}^3 m_{2}^2}{r_{12}^4} + \frac{92}{15} \frac{G^3 m_{1}^2
  m_{2}^2}{r_{12}^3} (n_{12}{} v_{1}{})^2\nonumber\\ &\quad +
\frac{G^3 m_{1}^3 m_{2}}{r_{12}^3} \Bigl(\frac{152}{15} (n_{12}{}
v_{1}{})^2 - \frac{56}{3} (n_{12}{} v_{1}{}) (n_{12}{} v_{2}{}) +
\frac{576}{35} (v_{1}{} v_{2}{}) - \frac{288}{35}
v_{1}{}^{2}\Bigr)\nonumber\\ &\quad + \frac{G^3 m_{1}
  m_{2}^3}{r_{12}^3} \Bigl(- \frac{176}{15} (n_{12}{} v_{1}{})^2 +
\frac{288}{35} v_{1}{}^{2}\Bigr)\nonumber\\ &\quad + \frac{G^3 m_{1}^2
  m_{2}^2}{r_{12}^4} \Bigl(- \frac{8}{35} (n_{12}{} v_{1}{})^2
(n_{12}{} y_{1}{}) + \frac{16}{35} (n_{12}{} v_{1}{}) (n_{12}{}
v_{2}{}) (n_{12}{} y_{1}{}) - \frac{8}{35} (n_{12}{} v_{2}{})^2
(n_{12}{} y_{1}{}) \nonumber\\ &\qquad\quad - \frac{64}{35} (v_{1}{}
v_{2}{}) (n_{12}{} y_{1}{}) - \frac{16}{21} (n_{12}{} v_{1}{})
(v_{1}{} y_{1}{}) + \frac{16}{21} (n_{12}{} v_{2}{}) (v_{1}{} y_{1}{})
+ \frac{16}{21} (n_{12}{} v_{1}{}) (v_{2}{} y_{1}{})
\nonumber\\ &\qquad\quad - \frac{16}{21} (n_{12}{} v_{2}{}) (v_{2}{}
y_{1}{}) + \frac{32}{35} (n_{12}{} y_{1}{}) v_{1}{}^{2} +
\frac{32}{35} (n_{12}{} y_{2}{}) v_{1}{}^{2}\Bigr)\nonumber\\ &\quad +
\frac{G^2 m_{1} m_{2}^2}{r_{12}^2} \Bigl(8 (n_{12}{} v_{1}{})^4 - 32
(n_{12}{} v_{1}{})^3 (n_{12}{} v_{2}{}) + \frac{44}{5} (n_{12}{}
v_{1}{})^2 (v_{1}{} v_{2}{}) - \frac{22}{5} (n_{12}{} v_{1}{})^2
v_{1}{}^{2} \nonumber\\ &\qquad\quad + \frac{56}{5} (n_{12}{} v_{1}{})
(n_{12}{} v_{2}{}) v_{1}{}^{2} - \frac{28}{5} (n_{12}{} v_{2}{})^2
v_{1}{}^{2} + \frac{584}{105} (v_{1}{} v_{2}{}) v_{1}{}^{2} -
\frac{188}{105} v_{1}{}^{4}\Bigr)\nonumber\\ &\quad + \frac{G^2
  m_{1}^2 m_{2}}{r_{12}^2} \Bigl(-8 (n_{12}{} v_{1}{})^4 + 32
(n_{12}{} v_{1}{})^3 (n_{12}{} v_{2}{}) - 48 (n_{12}{} v_{1}{})^2
(n_{12}{} v_{2}{})^2 - \frac{56}{5} (n_{12}{} v_{1}{})^2 (v_{1}{}
v_{2}{}) \nonumber\\ &\qquad\quad + \frac{112}{5} (n_{12}{} v_{1}{})
(n_{12}{} v_{2}{}) (v_{1}{} v_{2}{}) + \frac{100}{21} (v_{1}{}
v_{2}{})^2 + \frac{28}{5} (n_{12}{} v_{1}{})^2 v_{1}{}^{2} -
\frac{56}{5} (n_{12}{} v_{1}{}) (n_{12}{} v_{2}{}) v_{1}{}^{2}
\nonumber\\ &\qquad\quad + \frac{22}{5} (n_{12}{} v_{2}{})^2
v_{1}{}^{2} - \frac{584}{105} (v_{1}{} v_{2}{}) v_{1}{}^{2} +
\frac{188}{105} v_{1}{}^{4} + \frac{292}{105} v_{1}{}^{2}
v_{2}{}^{2}\Bigr)\biggl\}\bm{n}_{12}\nonumber\\
%%%%%%%%%%%%%%%%%%%%%%%%%%%%%%%%%%%%%%%%%%%%%%%%%%%%%%%%%%%%%%%%%%
&\quad + \biggl \{\frac{G^3 m_{1}^2 m_{2}^2}{r_{12}^3} \Bigl(-
\frac{12}{5} (n_{12}{} v_{1}{}) - 4 (n_{12}{} v_{2}{})\Bigr) +
\frac{G^3 m_{1}^3 m_{2}}{r_{12}^3} \Bigl(- \frac{8}{15} (n_{12}{}
v_{1}{}) + \frac{8}{15} (n_{12}{} v_{2}{})\Bigr)\nonumber\\ &\quad +
\frac{G^3 m_{1} m_{2}^3}{r_{12}^3} \Bigl(\frac{8}{15} (n_{12}{}
v_{1}{}) + \frac{16}{15} (n_{12}{} v_{2}{})\Bigr)\nonumber\\ &\quad +
\frac{G^3 m_{1}^2 m_{2}^2}{r_{12}^4} \Bigl(- \frac{16}{21} (n_{12}{}
v_{1}{}) (n_{12}{} y_{1}{}) + \frac{16}{21} (n_{12}{} v_{2}{})
(n_{12}{} y_{1}{}) + \frac{32}{35} (v_{1}{} y_{1}{}) - \frac{32}{35}
(v_{2}{} y_{1}{}) \nonumber\\ &\qquad\quad - \frac{16}{21} (n_{12}{}
v_{1}{}) (n_{12}{} y_{2}{}) + \frac{16}{21} (n_{12}{} v_{2}{})
(n_{12}{} y_{2}{}) + \frac{32}{35} (v_{1}{} y_{2}{}) - \frac{32}{35}
(v_{2}{} y_{2}{})\Bigr)\nonumber\\ &\quad + \frac{G^2 m_{1}
  m_{2}^2}{r_{12}^2} \Bigl(- \frac{52}{5} (n_{12}{} v_{1}{})^3 +
\frac{156}{5} (n_{12}{} v_{1}{})^2 (n_{12}{} v_{2}{}) - \frac{156}{5}
(n_{12}{} v_{1}{}) (n_{12}{} v_{2}{})^2 + \frac{52}{5} (n_{12}{}
v_{2}{})^3 \nonumber\\ &\qquad\quad - \frac{304}{15} (n_{12}{}
v_{1}{}) (v_{1}{} v_{2}{}) + \frac{328}{15} (n_{12}{} v_{2}{})
(v_{1}{} v_{2}{}) + \frac{152}{15} (n_{12}{} v_{1}{}) v_{1}{}^{2} -
\frac{164}{15} (n_{12}{} v_{2}{}) v_{1}{}^{2} \nonumber\\ &\qquad\quad
+ \frac{152}{15} (n_{12}{} v_{1}{}) v_{2}{}^{2} - \frac{164}{15}
(n_{12}{} v_{2}{}) v_{2}{}^{2}\Bigr)\nonumber\\ &\quad + \frac{G^2
  m_{1}^2 m_{2}}{r_{12}^2} \Bigl(\frac{52}{5} (n_{12}{} v_{1}{})^3 -
\frac{156}{5} (n_{12}{} v_{1}{})^2 (n_{12}{} v_{2}{}) + \frac{156}{5}
(n_{12}{} v_{1}{}) (n_{12}{} v_{2}{})^2 - \frac{52}{5} (n_{12}{}
v_{2}{})^3 \nonumber\\ &\qquad\quad + \frac{304}{15} (n_{12}{}
v_{1}{}) (v_{1}{} v_{2}{}) - \frac{304}{15} (n_{12}{} v_{2}{})
(v_{1}{} v_{2}{}) - \frac{152}{15} (n_{12}{} v_{1}{}) v_{1}{}^{2} +
\frac{152}{15} (n_{12}{} v_{2}{}) v_{1}{}^{2} \nonumber\\ &\qquad\quad
- \frac{152}{15} (n_{12}{} v_{1}{}) v_{2}{}^{2} + \frac{152}{15}
(n_{12}{} v_{2}{}) v_{2}{}^{2}\Bigr)\biggl\}
\bm{v}_{1}\nonumber\\ &\quad + \frac{G^3 m_{1}^2 m_{2}^2}{r_{12}^4}
\Bigl(\frac{164}{105} (n_{12}{} v_{1}{})^2 - \frac{328}{105} (n_{12}{}
v_{1}{}) (n_{12}{} v_{2}{}) + \frac{164}{105} (n_{12}{} v_{2}{})^2 +
\frac{352}{105} (v_{1}{} v_{2}{}) \nonumber\\ &\qquad\quad -
\frac{176}{105} v_{1}{}^{2} - \frac{176}{105} v_{2}{}^{2}\Bigr)
\bm{y}_{1} + 1\leftrightarrow 2\,.
\end{align}
\end{subequations}
and
\begin{subequations}\label{schottIharm}
\begin{align}
\mathbf{G}^\mathrm{2.5PN} &= \frac{4G m_1 m_2}{5c^5}\left(v_{12}^2 -
\frac{2G (m_1+m_2)}{r_{12}}\right)\bm{v}_{1} + 1\leftrightarrow 2\,,\\
%%%%%%%%%%%%%%%%%%%%%%%%%%%%%%%%%%%%%%%%%%%%%%%%%%%%%%%%%%%%%
\mathbf{G}^\mathrm{3.5PN} &= \biggl \{- \frac{8}{3} \frac{G^3 m_{1}^3
  m_{2}}{r_{12}^2} (n_{12}{} v_{1}{}) + \frac{24}{35} \frac{G^3
  m_{1}^2 m_{2}^2}{r_{12}^2} (n_{12}{} v_{1}{}) + \frac{52}{15}
\frac{G^3 m_{1} m_{2}^3}{r_{12}^2} (n_{12}{}
v_{1}{})\nonumber\\ &\quad + \frac{G^2 m_{1} m_{2}^2}{r_{12}} \Bigl(-
\frac{8}{5} (n_{12}{} v_{1}{})^3 + \frac{24}{5} (n_{12}{} v_{1}{})^2
(n_{12}{} v_{2}{}) + \frac{4}{15} (n_{12}{} v_{1}{}) (v_{1}{} v_{2}{})
\nonumber\\ &\qquad\quad - \frac{2}{15} (n_{12}{} v_{1}{}) v_{1}{}^{2}
- \frac{4}{15} (n_{12}{} v_{2}{}) v_{1}{}^{2}\Bigr)\nonumber\\ &\quad
+ \frac{G^2 m_{1}^2 m_{2}}{r_{12}} \Bigl(\frac{8}{5} (n_{12}{}
v_{1}{})^3 - \frac{24}{5} (n_{12}{} v_{1}{})^2 (n_{12}{} v_{2}{}) +
\frac{8}{15} (n_{12}{} v_{1}{}) (v_{1}{} v_{2}{})
\nonumber\\ &\qquad\quad - \frac{76}{105} (n_{12}{} v_{1}{})
v_{1}{}^{2} - \frac{2}{15} (n_{12}{} v_{2}{})
v_{1}{}^{2}\Bigr)\biggl\} \bm{n}_{12} \nonumber\\
%%%%%%%%%%%%%%%%%%%%%%%%%%%%%%%%%%%%%%%%%%%%%%%%%%%%%%%%%%%%%%
&\quad + \biggl \{- \frac{4}{15} \frac{G^3 m_{1}^2 m_{2}^2}{r_{12}^2}
- \frac{172}{105} \frac{G^3 m_{1} m_{2}^3}{r_{12}^2} + \frac{34}{21}
\frac{G^2 m_{1}^3}{r_{12}} v_{1}{}^{2}\nonumber\\ &\quad + \frac{G^2
  m_{1} m_{2}^2}{r_{12}} \Bigl(\frac{44}{15} (n_{12}{} v_{1}{})^2 -
\frac{88}{15} (n_{12}{} v_{1}{}) (n_{12}{} v_{2}{}) + \frac{56}{15}
(n_{12}{} v_{2}{})^2 \nonumber\\ &\qquad\quad - \frac{68}{35} (v_{1}{}
v_{2}{}) + \frac{6}{35} v_{1}{}^{2} + \frac{188}{105}
v_{2}{}^{2}\Bigr)\nonumber\\ &\quad + \frac{G^2 m_{1}^2 m_{2}}{r_{12}}
\Bigl(- \frac{104}{105} (n_{12}{} v_{1}{})^2 + \frac{88}{15} (n_{12}{}
v_{1}{}) (n_{12}{} v_{2}{}) - \frac{44}{15} (n_{12}{} v_{2}{})^2
\nonumber\\ &\qquad\quad - \frac{1144}{105} (v_{1}{} v_{2}{}) +
\frac{634}{105} v_{1}{}^{2} + \frac{118}{15}
v_{2}{}^{2}\Bigr)\nonumber\\ &\quad + G m_{1}^2 \Bigl(\frac{34}{21}
(v_{1}{} v_{2}{}) v_{1}{}^{2} - \frac{13}{15} v_{1}{}^{4} -
\frac{17}{21} v_{1}{}^{2} v_{2}{}^{2}\Bigr)\nonumber\\ &\quad + G
m_{1} m_{2} \Bigl(- \frac{4}{7} (v_{1}{} v_{2}{})^2 - \frac{36}{35}
(v_{1}{} v_{2}{}) v_{1}{}^{2} + \frac{6}{7} v_{1}{}^{4} +
\frac{22}{21} (v_{1}{} v_{2}{}) v_{2}{}^{2} \nonumber\\ &\qquad\quad +
\frac{11}{105} v_{1}{}^{2} v_{2}{}^{2} - \frac{7}{15}
v_{2}{}^{4}\Bigr)\biggl\} \bm{v}_{1} \nonumber\\
%%%%%%%%%%%%%%%%%%%%%%%%%%%%%%%%%%%%%%%%%%%%%%%%%%%%%%%%%%%%%%
&\quad + \biggl \{- \frac{4}{5} \frac{G^3 m_{1}^3 m_{2}}{r_{12}^3}
(n_{12}{} v_{1}{}) + \frac{G^3 m_{1}^2 m_{2}^2}{r_{12}^3}
\Bigl(\frac{4}{5} (n_{12}{} v_{1}{}) - \frac{4}{5} (n_{12}{}
v_{2}{})\Bigr) + \frac{4}{5} \frac{G^3 m_{1} m_{2}^3}{r_{12}^3}
(n_{12}{} v_{2}{})\nonumber\\ &\quad + \frac{G^2 m_{1}^2
  m_{2}}{r_{12}^2} \Bigl(- \frac{4}{5} (n_{12}{} v_{1}{}) (v_{1}{}
v_{2}{}) + \frac{2}{5} (n_{12}{} v_{1}{}) v_{1}{}^{2} + \frac{2}{5}
(n_{12}{} v_{1}{}) v_{2}{}^{2}\Bigr)\nonumber\\ &\quad + \frac{G^2
  m_{1} m_{2}^2}{r_{12}^2} \Bigl(\frac{4}{5} (n_{12}{} v_{2}{})
(v_{1}{} v_{2}{}) - \frac{2}{5} (n_{12}{} v_{2}{}) v_{1}{}^{2} -
\frac{2}{5} (n_{12}{} v_{2}{}) v_{2}{}^{2}\Bigr)\biggl\} \bm{y}_{1} +
1\leftrightarrow 2\,.
\end{align}
\end{subequations}
As we said, showing the existence of these 2.5PN and 3.5PN
contributions to the linear momentum and center of mass constitutes a
full proof of the balance equations~\eqref{dPidt}
and~\eqref{dGidt}. Note that there is no point about trying to relate
theses terms to those, given by Eqs.~\eqref{schottPi}
and~\eqref{schottIi}, obtained in the more general investigation of
Sec.~\ref{sec:balance}. Indeed, the calculation done in
Sec.~\ref{sec:balance} used the extended Burke-Thorne coordinate
system~\cite{B97} while the present calculation employs harmonic
coordinates.

We pospone to further work the discussion of the radiation-reaction terms in
the energy $E$ and angular momentum $J_i$ in a general frame. Note that these
have already been given in the frame of the center of mass by Iyer \&
Will~\cite{IW93,IW95} and, in the harmonic gauge, by Ref.~\cite{BBFM17}; see
Eqs.~(6.7) and (6.9) there.

\section{Discussion and conclusion} 
\label{sec:disc}

We have obtained the equations for the secular evolution by GW
emission of the linear momentum $\bm{P}$ and center-of-mass position
$\bm{G}$ of an isolated post-Newtonian source,
\begin{subequations}\label{fluxeqs}
\begin{align}
\frac{\ud \bm{P}}{\ud t} &= - \bm{F}_P\,,\\ \frac{\ud \bm{G}}{\ud t}
&= \bm{P} - \bm{F}_G\,,
\end{align}
\end{subequations}
where $\bm{F}_P$ and $\bm{F}_G$ represent the fluxes given
by~~\eqref{balanceP} and~\eqref{balanceG}, or more generally
by~\eqref{FluxP} and~\eqref{FluxG}. Let us now consider the case, to
start with, where the source is stationary before some instant $t_0$,
then emits a pulse of gravitational waves with finite duration between
times $t_0$ and $t_1$, and finally comes back to a stationary state at
later times $t>t_1$. This means that the fluxes $\bm{F}_P$ and
$\bm{F}_G$ are zero outside the time of emission, when $t<t_0$ and
$t>t_1$.

In this situation, it is straightforward to find the form of the
solution to Eqs.~\eqref{fluxeqs}. Initially, the linear momentum is
constant, so that, by applying a Lorentz boost, we can put ourselves
in the rest frame of the source, thus achieving $\bm{P}_0=\bm{0}$ (for
$t<t_0$). Furthermore, we can translate the origin of our coordinate
system in such a way that it coincides with the center of mass of the
source, hence $\bm{G}_0=\bm{0}$ initially. Then, by
integrating~\eqref{fluxeqs}, we get (for $t_0<t<t_1$)
\begin{subequations}\label{GWem}
\begin{align}
\bm{P}(t) &= - \int_{t_0}^{t} \ud t' \,\bm{F}_P(t')\,,\\ \bm{G}(t) &=
- \int_{t_0}^{t} \ud t' \,(t-t')\,\bm{F}_P(t') - \int_{t_0}^{t} \ud t'
\,\bm{F}_G(t')\,.
\end{align}
\end{subequations}
After the period of emission (for $t>t_1$), the source is again
stationary but has acquired a net constant linear momentum $\bm{P}_1$
with respect to its initial value $\bm{P}_0=\bm{0}$ and the motion of
its center of mass has become uniform, \textit{i.e.}
$\bm{G}_1=\bm{P}_1\,t+\bm{Z}_1$. We find
\begin{subequations}\label{afterGW}
\begin{align}
\bm{P}_1 &= - \int_{t_0}^{t_1} \ud t' \,\bm{F}_P(t')\,,\\ \bm{Z}_1 &=
\int_{t_0}^{t_1} \ud t' \Bigl[ t'\,\bm{F}_P(t') -
  \bm{F}_G(t')\Bigr]\,.
\end{align}
\end{subequations}
As we see, the cumulative effect of the flux $\bm{F}_G$ results in
adding an extra contribution to the position of the center of mass
after the GW emission. On the other hand, the final value of the
linear momentum $\bm{P}_1$ provides the total gravitational recoil
velocity of the source (or kick), as measured in the asymptotic
Minkowskian frame:
\begin{equation}\label{V1}
\bm{V}_1 = \frac{\bm{P}_1}{\sqrt{M_1^2+\frac{\bm{P}_1^2}{c^2}}}\,,
\end{equation}
where $M_1$ is the final mass of the system, after the GW emission has
stopped. The variation of the mass is obtained by integrating the
energy balance equation,
\begin{equation}\label{M1M0}
M_1 = M_0 - \frac{1}{c^2}\int_{t_0}^{t_1} \ud t' \,F_E(t')\,,
\end{equation}
where $F_E$ is the energy flux, for instance given by~\eqref{balanceE}.
However, since the fluxes represent small quantities in the adiabatic
approximation, the mass $M_1$ under the square root of Eq.~\eqref{V1} might be
approximated by the initial value $M_0$,\footnote{Numerical relativity (NR)
  calculations show that the energy radiated away by black-hole binary systems
  represents a few percents of their total mass energy; see, \textit{e.g.},
  Fig.~9 of Ref.~\cite{GHS18} (top right panel).} whereas the relativistic
correction $\bm{P}_1^2/c^2$ can be neglected. The gravitational
recoil~\eqref{V1} has been investigated for general systems
in~\cite{BoR61,Peres62,Papa71,Bek73,Th80} and some estimations have been
proposed for binary systems in~\cite{Fit83,Wi92,K95,BQW05,RBK09,LBW10}.

We consider next the ``\textit{instantaneous}'' motion of the center
of mass during the GW emission, as given by Eqs.~\eqref{GWem}. To be
specific, we focus on the case of a Newtonian binary system with no
spins and moving on an exactly circular orbit (we neglect the
radiation reaction on the orbit). We introduce the symmetric mass
ratio $\nu=m_1m_2/m^2$ with $m=m_1+m_2$, assuming $m_1\geqslant m_2$
so that $m_1-m_2=m\sqrt{1-4\nu}$. The orbital separation is denoted by
$r$ and the unit vector along the binary's separation (pointing
towards the larger mass $m_1$) by $\bm{n}$. We also define the unit
vector $\bm{\lambda}$ orthogonal to $\bm{n}$ in the orbital plane and
oriented in the sense of the relative motion. For circular orbits, the
relative velocity of the particles reduces to
$\bm{v}=r\omega\bm{\lambda}$, where $\omega=\sqrt{G m/r^3}$ is the
(Newtonian) orbital frequency.

The RHS of the linear momentum flux equation~\eqref{dPidt} is
straightforward to evaluate for the Newtonian circular binary system
with result~\cite{Fit83}
\begin{equation}\label{dPdtbin}
\frac{\ud \bm{P}}{\ud t} =
\frac{464}{105}\,\frac{G^4m^5\omega}{c^7r^4}\,\sqrt{1-4\nu}\,\nu^2\,
\bm{\lambda}\,.
\end{equation}
This relation holds at any time along the orbit and can be integrated,
yielding
\begin{equation}\label{Pbin}
\bm{P} = \frac{464}{105}\,\frac{G^4m^5}{c^7r^4}\,\sqrt{1-4\nu}\,\nu^2\,\bm{n}\,,
\end{equation}
where we assume from now on that a Lorentz boost and a shift of the
origin of the coordinate system have been applied to set $\bm{P}$ and
$\bm{G}$ to zero when averaged over an orbit (neglecting the
radiation-reaction decay). Then, we use the center-of-mass balance
equation~\eqref{dGidt}, which leads to
\begin{equation}\label{dGdtbin}
\frac{\ud \bm{G}}{\ud t} = \bm{P} +
\frac{544}{105}\,\frac{G^4m^5}{c^7r^4}\,\sqrt{1-4\nu}\,\nu^2\,\bm{n}
\,,
\end{equation}
so that, combining the two previous results and integrating,
\begin{equation}\label{Gbin}
\bm{G} = -
\frac{48}{5}\,\frac{G^4m^5}{c^7r^4\omega}\,\sqrt{1-4\nu}\,\nu^2\,\bm{\lambda}
\,.
\end{equation}
The equations~\eqref{Pbin} and~\eqref{Gbin} give the instantaneous values of
the momentum and center-of-mass position of a circular binary system
(neglecting the orbital decay). Their RHS are equal to minus the values that
can be attributed to the gravitational radiation field. It would be
interesting to compare the prediction~\eqref{Gbin} for the oscillations of the
center of mass with very accurate numerical computations of the recoil and
center-of-mass position, such as those of Ref.~\cite{GHS18}. Perhaps the problem, in
performing such comparison, would be the control of the different gauges used by PN
and NR calculations.

To conclude, we have argued that the usual ``quadrupole-type'' formulas for
the energy, angular momentum and linear momentum, are missing an analogous
formula for the position of the center of mass. Indeed, the complete set of
invariants of a relativistic system does include the (initial position of the)
center of mass, which is the Noetherian quantity associated with the
invariance of the dynamics under Lorentz boosts. We have found three
derivations of the balance equation describing the effect of GW emission on
the position of the center of mass. The first one is based on the local
equations of motion of a general isolated post-Newtonian source, including the
gravitational radiation-reaction force at the 3.5PN order in a specific gauge,
which have been integrated over the volume of the source, leading to the
requested flux equation [see Secs.~\ref{sec:radreac} and~\ref{sec:balance}].
The second derivation, also valid for a general isolated source, is a direct
flux calculation performed at future null infinity, yielding the full
multipole moment expansion for the flux, though restricted to the dominant
post-Minkowskian order [see Sec.~\ref{sec:flux}]. Finally, we have verified
that, for the known radiation-reaction force at the 3.5PN order in the
particular case of compact binary systems, the balance equation is indeed
satisfied [see Sec.~\ref{sec:eom}]. Independent and complementary derivations
of the balance equation for the center-of-mass position have been achieved
recently~\cite{KNQ18,N18}. Further work is needed to know whether there could
be any possible interest and/or application for this effect in Astrophysics.

\acknowledgments We are grateful to Clifford Will and Alexandre Le Tiec for
interesting discussions. Moreover, we appreciated the comments made on a first version
of this paper by Bala Iyer, Carlos Kozameh, David Nichols and Leo Stein.

\appendix

\bibliography{BF18decsub.bib}

\end{document}